\documentclass[apj,useAMS,usenatbib,usegraphicx,upmath]{emulateapj} \usepackage{amsmath} \usepackage{amsfonts} \usepackage{amssymb}  \usepackage{apjfonts}\usepackage{color}

\pdfoutput = 1

\shorttitle{Electrodynamics}
\shortauthors{T.T. Koskinen et al.}

\begin{document}

\title{Electrodynamics on extrasolar giant planets}

\author{T. T. Koskinen\altaffilmark{1}, R. V. Yelle\altaffilmark{1}, P. Lavvas\altaffilmark{2}, J. Y-K. Cho\altaffilmark{3,4}}
\altaffiltext{1}{Lunar and Planetary Laboratory, University of Arizona, 1629 E. University Blvd., Tucson, AZ 85721--0092; tommi@lpl.arizona.edu}
\altaffiltext{2}{Groupe de Spectroscopie Mol\'eculaire et Atmosph\'erique UMR CNRS 7331, Universit\'e Reims Champagne-Ardenne, 51687 Reims, France}
\altaffiltext{3}{Astronomy Unit, School of Mathematical Sciences, Queen Mary, University of London, Mile End Road, London E1 4NS, UK}
\altaffiltext{4}{Institute for Theory and Computation, Harvard University, Cambridge, MA 02138}

\begin{abstract}

Strong ionization on close-in extrasolar giant planets suggests that their atmospheres may be affected by ion drag and resistive heating arising from wind-driven electrodynamics.  Recent models of ion drag on these planets, however, are based on thermal ionization only and do not include the upper atmosphere above the 1 mbar level.  These models are also based on simplified equations of resistive MHD that are not always valid in extrasolar planet atmospheres.  We show that photoionization dominates over thermal ionization over much of the dayside atmosphere above the 100 mbar level, creating an upper ionosphere dominated by ionization of H and He and a lower ionosphere dominated by ionization of metals such as Na, K, and Mg.  The resulting dayside electron densities on close-in exoplanets are higher than those encountered in any planetary ionosphere of the solar system, and the conductivities are comparable to the chromosphere of the Sun.  Based on these results and assumed magnetic fields, we constrain the conductivity regimes on close-in EGPs and use a generalized Ohm's law to study the basic effects of electrodynamics in their atmospheres.  We find that ion drag is important above the 10 mbar level where it can also significantly alter the energy balance through resistive heating.  Due to frequent collisions of the electrons and ions with the neutral atmosphere, however, ion drag is largely negligible in the lower atmosphere below the 10 mbar level for a reasonable range of planetary magnetic moments.  We find that the atmospheric conductivity decreases by several orders of magnitude in the night side of tidally locked planets, leading to a potentially interesting large scale dichotomy in electrodynamics between the day and night sides.  A combined approach that relies on UV observations of the upper atmosphere, phase curve and Doppler measurements of global dynamics, and visual transit observations to probe the alkali metals can potentially be used to constrain electrodynamics in the future.                
               
\end{abstract}


\section{Introduction}
\label{sc:intro}   

The discovery of thousands of extrasolar planet systems \citep[e.g.,][]{udry07,batalha13,tenenbaum14} and the ongoing efforts to characterize the atmospheres on many of these planets greatly expand the scope of atmospheric science from the limited sample of planets in the solar system.  New regimes of thermal structure, dynamics, and escape have all been subject to intense scrutiny, especially on close-in extrasolar giant planets (EGPs) or Hot Jupiters that typically orbit within 0.1 AU of their host stars, the objects for which we currently have the best observational constraints.  More recently, ionization of these atmospheres has been studied because of the recognition that it may strongly affect the temperatures and dynamics of hot exoplanet atmospheres \citep[e.g.,][]{cho08,batygin10,perna10a,perna10b,rogers14}.  These studies have employed a number of simplifying assumptions, primarily using the induction equation for resistive MHD, and reached often contradictory conclusions.  There is, as yet, no general agreement on the importance of ion drag or resistive heating in exoplanet atmospheres.

Our study here expands on the previous analyses by considering photoionization as well as thermal ionization of the atmosphere.  We find that photoionization dominates on the dayside and is responsible for the primary effects of ion drag on the neutral atmosphere.  We also examine in detail the role of collisions and anisotropic resistivity, effects usually neglected or treated approximately in resistive MHD, and show that ion drag strongly couples different atmospheric regions and may dominate the dynamics of the middle and upper atmosphere of close-in EGPs\footnote{In this work we define the lower atmosphere to be below the 0.01 bar level and the middle atmosphere to be between 0.01 bar and 10$^{-6}$ bar.}.  Because our knowledge of exoplanet atmospheric structure and dynamics is still rudimentary, detailed predictions of electrodynamics on EGPs are not yet possible.  Nevertheless, our results show that ion drag may dominate the dynamics and temperature structure in some regions of the atmosphere and therefore cannot be neglected.

Transit observations have revealed the presence of escaping ions and neutral atoms in the upper atmospheres of close-in EGPs such as HD209458b, HD189733b, and WASP-12b \citep[e.g.,][]{vidalmadjar03,linsky10,lecavelier12,benjaffel13,fossati10}.  As a result, previous studies of photoionization have mostly concentrated on the thermosphere ($p \lesssim$~10$^{-6}$ bar) where its role in ionizing hydrogen and helium, heating the atmosphere and powering mass loss is well recognized \citep[e.g.,][]{yelle04,garciamunoz07,koskinen13a,koskinen13b}.  The effective temperatures of many close-in EGPs, however, are high enough for alkali metals such as Na and K to remain in the atmosphere as atoms instead of condensing or forming molecules.  

Both Na and K have been detected on different Hot Jupiters \citep[e.g.,][]{sing08a,sing11,lavvas14} and the detection of Na on the well known transiting planet HD209458b constituted the first detection of an exoplanet atmosphere \citep{charbonneau02}.  The ionization potentials of these metals are relatively low and thus they can be effectively ionized both thermally and by photoionization throughout the atmosphere.  As we will demonstrate, this leads to higher electron densities in the middle and lower atmospheres than in the thermosphere.  As a results, the conductivities in close-in EGP atmospheres are higher than in any planetary ionosphere of the solar system, and in fact closer to the conductivities in the outer atmospheres of stars like the Sun.     

We use a photochemical model to calculate the thermal ionization and photoionization rates between 100 bar and 10$^{-10}$ bar, and identify the basic conductivity regimes in close-in EGP atmospheres based on the resulting electron densities and assumed magnetic field strengths.  The photochemical calculations are based on a prescribed temperature profile and properly include the shielding of the relevant metals from ionizing radiation by other atoms and molecules.  The details of these simulations are presented in a companion paper \citep{lavvas14}.  We use the results to constrain the composition of the ionosphere and, together with a generalized Ohm's law, to study the basic mechanisms of electrodynamics in hot EGP atmospheres.  

We treat HD209458b ($R_p =$~1.32 $R_{\text{Jup}}$, $M_p =$~0.69 $M_{\text{Jup}}$, $a =$~0.047 AU)\footnote{See www.exoplanet.eu for more details} as a prototypical Hot Jupiter and use its properties in all of the calculations below.  The general results of this study, however, are not limited to any specific Hot Jupiter.  As usual, we also assume that the rotation rate of HD209458b is consistent with tidal locking, given that the synchronization timescale of its rotation period is much shorter than the age of the system \citep{guillot96}.  The magnetic fields of EGPs are currently unconstrained by observations, neither do we possess a robust theory for planetary dynamos with definite predictive capabilities.  Given these uncertainties, we assume a basic untilted dipole field with the Jovian magnetic moment of $\mu_J =$~1.56~$\times$~10$^{27}$ A~m$^2$ in all calculations below, unless otherwise explicitly indicated.                
          
\section{Basic properties of the ionosphere}
\label{sc:methods}     

The importance of electrodynamics and ion drag varies greatly between different regions of the atmosphere.  In addition to the plasma density, it depends on the magnetization of the plasma that affects the conductivities.  Magnetization is the coupling of the electrons and ions to the magnetic field in the atmosphere that is quantified by\footnote{Note that this parameter is also called the Hall parameter \citep{wardle07}.}:
\begin{equation}
k_{st} = \frac{\omega_s}{\nu_{st}}
\label{eqn:kst1}
\end{equation}  
where $\omega_s = \left| q_s \right| B/m_s$ is the gyrofrequency of species $s$ and $\nu_{st}$ is its collision frequency with other species $t$.  High magnetization of $k_s >> 1$, where $k_s = \sum_t k_{st}$, implies strong coupling to the magnetic field whereas low values of $k_s < 1$ indicate that collisions dominate.  

\subsection{Electron densities}
\label{subsc:ionosphere}

\begin{figure}
  \epsscale{1.15}
  \plotone{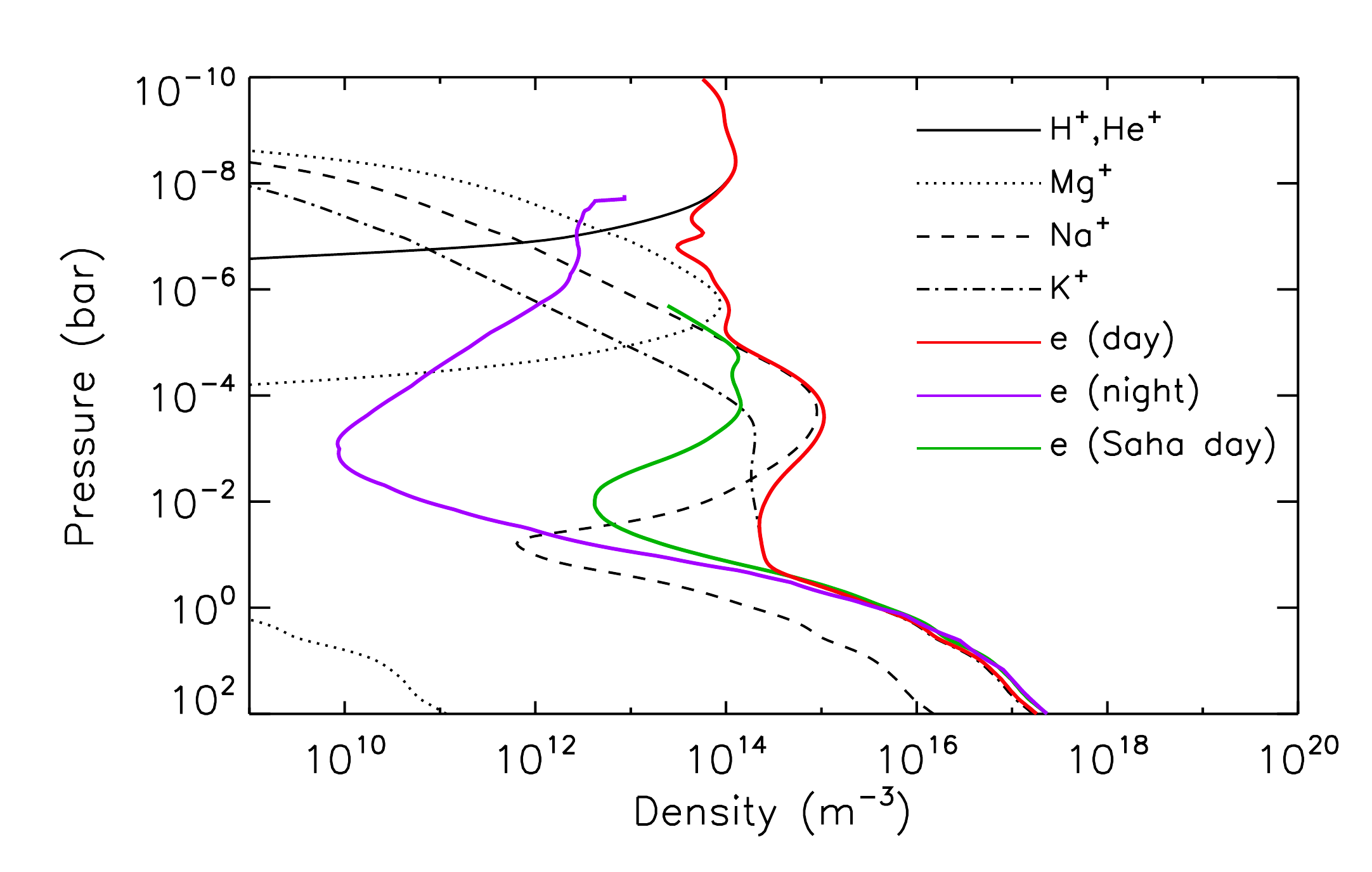}
  \caption{The ionosphere of HD209458b \citep{lavvas14}.  Black lines show the dayside ion densities, the red line shows the total dayside electron density, the green line shows the dayside electron density based on a simplified Saha equation \citep{menou12}, and the purple line shows the estimated night side electron density (see text).  Note that the densities of the heavy ions decrease rapidly at the lowest pressures because drag due to the escaping hydrogen \citep{koskinen13a,koskinen13b} is not included in the above model.}
  \label{fig:ionosphere}
\end{figure}

Electron densities on close-in EGPs are much higher than the electron densities in the ionospheres of the solar system.  To illustrate this, Figure~\ref{fig:ionosphere} shows the dominant ion and electron densities in the dayside ionosphere of HD209458b based on our model \citep{lavvas14}.  We used the same T-P profiles as \citet{moses11} to calculate the recombination and thermal ionization rates, and included neutral photochemistry and photoionization in the model.  Previous models of ion drag on close-in EGPs have only considered thermal ionization of the atmosphere \citep[e.g.,][]{perna10a,perna10b,batygin10}.  As shown by Figure~\ref{fig:ionosphere}, photoionization of abundant metals such as sodium, potassium and magnesium produce electron densities above the 1 bar level that are 10--100 times higher than the electron densities based on the Saha equation \citep{menou12}.  This means that the upper atmosphere above the 1--10 mbar level cannot be treated as an insulator, as suggested by \citet{batygin10}.

Naturally, photoionization does not take place in the night side.  Some of the electrons are, however, transported from the dayside to the night side by circulation even on tidally locked planets.  In addition to thermal ionization based on the night side T-P profile, we estimated the night side electron densities from a simple balance of advection and recombination between the dusk terminator and the anti-stellar point along the equator:   
\begin{eqnarray}
& &\frac{u_{\phi}}{r} \frac{\partial n_e}{\partial \phi} = - n_e^2 \Phi_{\text{eff}} \\
&\rightarrow& f_{dn} = \frac{n_{e\text{day}}}{n_{e\text{night}}} \approx 1 + \frac{r n_{e\text{day}} \Phi_{\text{eff}}}{u_{\phi}} \delta \phi 
\end{eqnarray}   
where $u_{\phi} =$1 km~s$^{-1}$ is the zonal wind speed \citep[e.g.,][]{showman02}\footnote{Wind speeds of the order of 1 km~s$^{-1}$ can be derived from scaling laws based on the thermal wind equation and plausible estimates of the energy balance.  They have also been predicted by numerous other circulation models.} and $\delta \phi = \pi/2$.  The effective recombination rate that is calculated by using the night side T-P profile is $\Phi_{\text{eff}} = (\sum_i n_i \Phi_i)/n_e$, where $\Phi_i$ are the recombination rates of the dominant ions and $(n_i/n_e)$ are the dayside ion fractions.  The night side electron densities are also shown in Figure~\ref{fig:ionosphere}.

Most of the dayside electrons above the 0.2 bar level are released by photoionization.  In line with previous results from \citet{koskinen10}, the electron density in the thermosphere ($p \lesssim$~10$^{-6}$ bar) is lower by about an order of magnitude in the night side than in the dayside.  Above the 10$^{-8}$ bar level, in the extended thermosphere, the line of sight optical depth is low enough to allow the stellar XUV radiation to penetrate to the night side and significant differences between day and night side electron densities are not expected.  Between 10$^{-6}$ bar and 0.2 bar, however, the dayside electron densities are orders of magnitude higher than the night side electron densities.  Below the 0.2 bar level, where thermal ionization dominates, there are again practically no diurnal differences.  

\subsection{Generalized Ohm's law}
\label{subsc:ohms}


Ion drag and resistive heating arise from electric currents in the atmosphere that are driven by perturbations to the magnetic field.  These perturbations, in turn, arise from variable electric fields induced by plasma motions.  The generalized Ohm's law provides a relationship between the currents and the electric fields that depends on atmospheric conductivities.  With Faraday's law, the Ohm's law can also be converted into an induction equation that describes the evolution of the magnetic field perturbations.  Here we outline the derivation of a generalized Ohm's law that is valid for both weakly and strongly ionized media and thus sufficiently flexible to be valid in most planetary atmospheres.  We use this Ohm's law in Section~\ref{sc:emdyn} to model electrodynamics.  We note that the generalized Ohm's law differs from the usual Ohm's law used in studies of planetary ionospheres in the solar system that typically assume weak ionization.  

The derivation is based on combining the ion, electron, and neutral momentum equations: 
\begin{eqnarray}
& &\rho_s \frac{\text{d} \mathbf{u}_s}{\text{d} t} + 2 \rho_s \boldsymbol\Omega \times \mathbf{u}_s + \nabla p_s - n_s q_s (\mathbf{E} + \mathbf{u}_s \times \mathbf{B}) - \rho_s \mathbf{g} \nonumber \\
& & = \sum_t \rho_s \nu_{st} (\mathbf{u}_t - \mathbf{u}_s),  
\label{eqn:momentum}
\end{eqnarray}
where $\boldsymbol\Omega$ is the angular velocity of planetary rotation, and eliminating the electron velocity by using the definition of current density:
\begin{equation}
\mathbf{j} = \sum_i q_i n_i \mathbf{u}_i - e n_e \mathbf{u}_e.
\end{equation}
The result for partly ionized media composed of a neutral species with mass $m_n$ and a single ion species with mass $m_i$ and charge $e$ is \citep[e.g.,][]{leake13}:  
\begin{eqnarray}
\mathbf{E}_i &=& \mathbf{E} + \mathbf{u}_i \times \mathbf{B} = \frac{B}{e n_p} \left[ \frac{1}{k_{ei}} + \frac{1}{k_{en}+k_{in}} \right] \mathbf{j} \nonumber \\ 
&+& \frac{B}{e n_p} \left[ 1 - Y_n \frac{k_{in} - (m_e/m_i) k_{en}}{k_{en}+k_{in}} \right] \mathbf{j} \times \mathbf{b}
\label{eqn:ion_ohm}
\end{eqnarray}  
where $n_p = n_e = n_i$ is the plasma density, $k_{st}$ is magnetization (for species $s$ colliding with $t$), $\mathbf{E}_i$ is the electric field in the rest frame of the ions, $Y_n = m_n n_n/(m_n n_n + m_i n_i)$ is the mass mixing ratio of the neutral atmosphere, and $\mathbf{b}$ is a unit vector in the direction of the magnetic field lines.  

In this work we ignore the inertia ($\text{d} \mathbf{u}_s/\text{d} t$), Coriolis, and pressure gradient terms in the Ohm's law.  The inertia terms can be neglected because the typical advection timescale $\tau_a \approx R_p/u \approx$~10$^5$ s, where $u \approx$~1 km~s$^{-1}$ is the characteristic wind speed on EGPs \citep[e.g.,][]{showman02}, is much longer than the plasma-neutral collision timescale of 10$^{-14}$--1 s.  By similar logic, the Coriolis force term is also negligible.  The pressure gradient terms can be neglected if      
\begin{equation}
\left( \frac{m_i}{m_n} \right) \beta_p  \left( \frac{c}{\Omega_i} \right) \frac{\mu_0 V_A}{\eta_P} << 1
\label{eqn:prescond}
\end{equation}  
where $c$ is the speed of light, $\mu_0$ is the permeability of free space, $\eta_P$ is the Pedersen resistivity (see below), $\beta_p$ is the plasma beta, $V_A$ is the plasma Alfven speed and $\Omega_i$ is the ion plasma frequency.  We find that inequality (\ref{eqn:prescond}) is true everywhere in our atmosphere for a planetary magnetic moment $\mu_p = \mu_J$.  When $\mu_p \lesssim 0.1 \mu_J$, however, inequality~(\ref{eqn:prescond}) does not hold in the upper atmosphere.   

It is convenient to write the Ohm's law in terms of the electric field in the center of mass frame $\mathbf{E}_{\text{cm}}$.  In this frame the velocity is given by:  
\begin{equation}
\mathbf{u}_{\text{cm}} \approx \mathbf{u}_i - Y_n \mathbf{w} - \frac{m_e}{m_i} \frac{Y_i}{e n_p} \mathbf{j}.
\end{equation}
where $\mathbf{w} = \mathbf{u}_i - \mathbf{u}_n$, which is obtained by combining the ion and neutral momentum equations:
\begin{equation}
\mathbf{w} = \left( \frac{k_{in}}{k_{en} + k_{in}} \right) \left[ \frac{\mathbf{j}}{e n_p}  + k_{en} \left( \frac{Y_n}{e n_p} \mathbf{j} \times \mathbf{b} \right) \right].
\end{equation}
Thus the Ohm's law in the center of mass frame is given by:
\begin{equation}
\mathbf{E}_{\text{cm}} = \mathbf{E} + \mathbf{u}_i \times \mathbf{B} - Y_n \mathbf{w} \times \mathbf{B} - Y_i \frac{1}{e n_p} \left( \frac{m_e}{m_i} \right) \mathbf{j} \times \mathbf{B}.
\end{equation}
The last term is negligible, and the result is 
\begin{equation}
\mathbf{E}_{\text{cm}} = \eta_{\parallel} \mathbf{j} - \eta_C (\mathbf{j} \times \mathbf{b}) \times \mathbf{b} + \eta_H \mathbf{j} \times \mathbf{b}
\label{eqn:ohm1}
\end{equation}
where the parallel ($\eta_{\parallel}$), Cowling ($\eta_C$) and Hall ($\eta_H$) resistivities are given by:
\begin{eqnarray}
\eta_{\parallel} &=& \frac{B}{e n_p} \left( \frac{1}{k_{ei}} + \frac{1}{k_{en} + k_{in}} \right) \\
\eta_C &=&  \frac{B}{e n_p} \left( \frac{Y_n^2 k_{en} k_{in}}{k_{en} + k_{in}} \right) \\
\eta_H &=&  \frac{B}{e n_p} \left[ \frac{k_{en} + k_{in} (1-2Y_n)}{k_{en} + k_{in}} \right]. 
\end{eqnarray} 

Taking the curl of equation~(\ref{eqn:ohm1}) and using Faraday's law yield the induction equation:
\begin{equation}
\frac{\partial \mathbf{B}}{\partial t} = \nabla \times \left[ \mathbf{u}_{\text{cm}} \times \mathbf{B} - \eta_{\parallel} \mathbf{j} + \eta_C (\mathbf{j} \times \mathbf{b}) \times \mathbf{b} - \eta_H \mathbf{j} \times \mathbf{b} \right].
\label{eqn:find1}
\end{equation}
Another way to write equation~(\ref{eqn:ohm1}) is:
\begin{equation}
\mathbf{j} = \overline{\sigma} \cdot \left( \mathbf{E}_{\text{cm}} + \mathbf{u}_{\text{cm}} \times \mathbf{B} \right)
\label{eqn:ohm2}
\end{equation}
where the conductivity tensor is given by:
\begin{eqnarray}
\overline{\sigma} &=& \left( \begin{array}{ccc}
\sigma_P & -\sigma_H & 0 \\
\sigma_H & \sigma_P & 0 \\
0 & 0 & \sigma_{\parallel} \end{array} \right).
\end{eqnarray}
The conductivities are related to the resistivities by: 
\begin{eqnarray}
\sigma_{\parallel} &=& \frac{1}{\eta_{\parallel}} \\
\sigma_P &=& \frac{\eta_P}{\eta_P^2 + \eta_H^2} \label{eqn:sigmap} \\
\sigma_H &=& \frac{\eta_H}{\eta_P^2 + \eta_H^2}
\end{eqnarray}
where $\eta_P = \eta_{\parallel} + \eta_C$.  Contrary to the fully ionized MHD equations that are used in many astrophysical applications or the weakly ionized approximations of ionospheric electrodynamics, equations~(\ref{eqn:find1}) and (\ref{eqn:ohm2}) are simultaneously valid for both cases, thus providing a useful connection between the two regimes.  We note that as $Y_n \rightarrow$~1, the conductivities in the center of mass frame are equivalent to the conductivities in the neutral frame that we will use in Section~\ref{sc:emdyn}.

\subsection{Basic plasma parameters}
\label{subsc:basics}

\begin{figure}
  \epsscale{1.15}
  \plotone{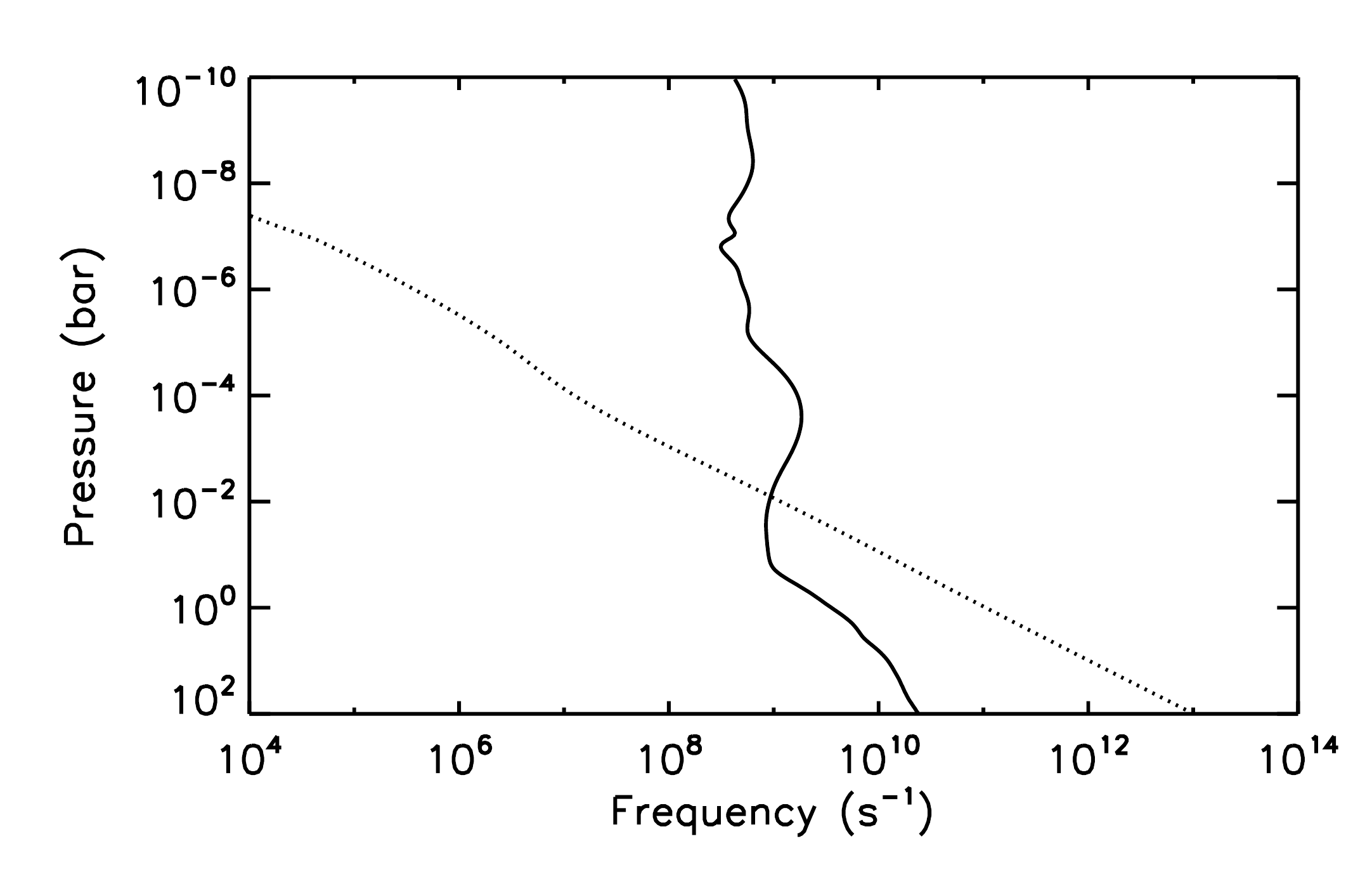}
  \caption{Plasma frequency ($\Omega_e$, solid line) and the electron-neutral collision frequency ($\nu_{en}$, dotted line) in the dayside ionosphere of HD209458b.}
  \label{fig:pfreqs}
\end{figure}

In addition to the conductivities and the Ohm's law, parameters such as the electron plasma frequency
\begin{equation}
\Omega_e = \sqrt{ \frac{n_e e^2}{\epsilon_0 m_e} }.
\end{equation}
provide basic insights to the dynamic regime in the ionospheres of close-in EGPs.  In Figure~\ref{fig:pfreqs} we compare the dayside plasma frequency with the electron-neutral collision frequency.  We have calculated the collision frequencies by including the neutrals H, H$_2$, and He, with rate expressions from \citet{koskinen10} and \citet{schunk00}.  The results indicate that $\nu_{en}$ exceeds $\Omega_e$ below the 0.01 bar level on the dayside and below the 10$^{-4}$ bar level in the night side, independently of the planetary magnetic field strength.  In this regime electron plasma waves are suppressed, the electrons (and ions) equilibrate with the neutral atmosphere and the medium does not behave as a plasma \citep[e.g.,][]{baumjohann97}.       

The Lundquist number measures the degree to which the background (planetary) magnetic field is perturbed by plasma motion, and it is given by \citep{leake13}:
\begin{equation}
S = \frac{\mu_0 V_{An}^2}{\eta_P \nu_{ni}}
\end{equation}
where 
$V_{An} = B / \sqrt{\mu_0 \rho}$ is the Alfven speed determined by using the total mass density $\rho = \rho_p + \rho_n$.  The Lundquist number is derived from the ratio of the advection term to the diffusion term in the basic induction equation:
\begin{equation}
\frac{\partial (\delta \mathbf{B})}{\partial t} = \nabla \times (\mathbf{u} \times \mathbf{B} - \frac{\eta}{\mu_0} \nabla \times \delta \mathbf{B}) 
\end{equation}
where $\delta \mathbf{B}$ is the perturbation to the background (planetary) magnetic field $\mathbf{B}$.  Contrary to the standard form of the magnetic Reynolds number $R_m$ \citep[e.g.,][]{baumjohann97}, the Lundquist number accounts for collisions with the neutral atmosphere and the stability of the planetary magnetic field that is generated in the interior of the planet.      

\begin{figure}
  \epsscale{1.15}
  \plotone{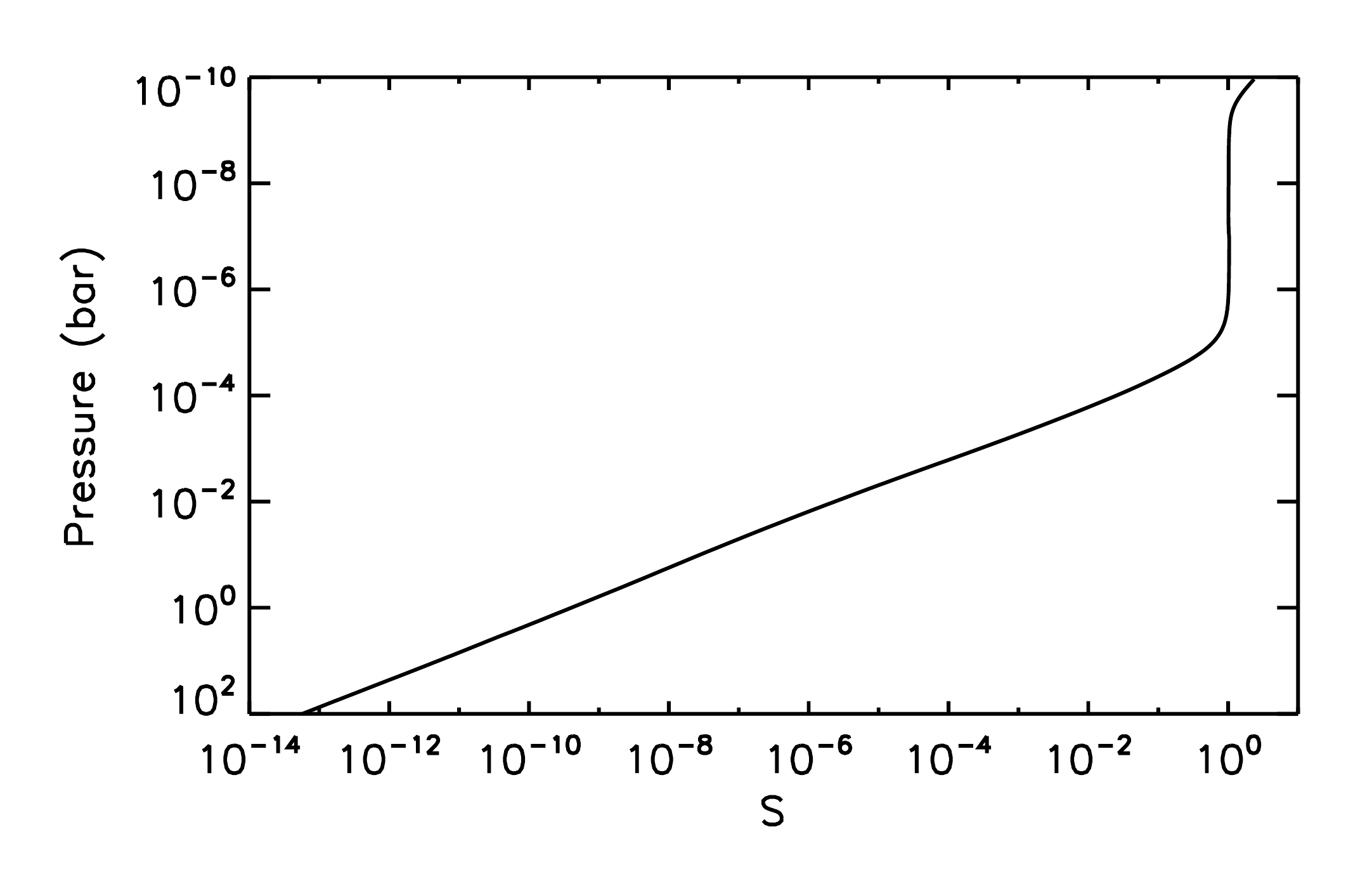}
  \caption{The Lundquist number $S$ in the dayside atmosphere of HD209458b.}
  \label{fig:lundq}
\end{figure}

Figure~\ref{fig:lundq} shows $S$ as a function of pressure in the dayside atmosphere of HD209458b.  The results indicate that $S$ reaches unity near the 10$^{-6}$--10$^{-5}$ bar level whereas in the lower atmosphere the values of $S$ are very low.  This is not a coincidence.  The Lundquist number is $S \approx k_{en} k_{in}$ below the 10$^{-3}$~bar level where the electron and ion magnetizations are both less than unity.  This suggests that significant perturbations to the planetary magnetic field are unlikely in the deep atmosphere, mostly because collisions with the neutral atmosphere suppress plasma behavior.  We note that this result is in contrast to the recent work by \citet{rogers14}.  In the upper and middle atmosphere, however, plasma dynamics can alter the planetary magnetic field and a solution to the full induction equation may be required to characterize the time evolution of the magnetic field.

\subsection{Conductivity and magnetization} 
\label{subsc:conds}  


The conductivities depend on the collision frequencies that also determine the degree of magnetization within the plasma.  Since the Ohm's law~(\ref{eqn:ohm2}) is given for a three fluid system and our ionosphere consists of several ions, we calculated the conductivities based on a mean ion species with the mass and collision frequencies given by:
\begin{equation}
\langle m_i \rangle = \frac{1}{n_p} \sum_i m_i n_i, \ \ \langle \nu_{in} \rangle = \frac{\sum_i m_i n_i \nu_{in}}{\sum_i m_i n_i}.
\end{equation}
We have found that the difference between this approach and fully accounting for several ion species in the Ohm's law is small. 

\begin{figure}
  \epsscale{1.15}
  \plotone{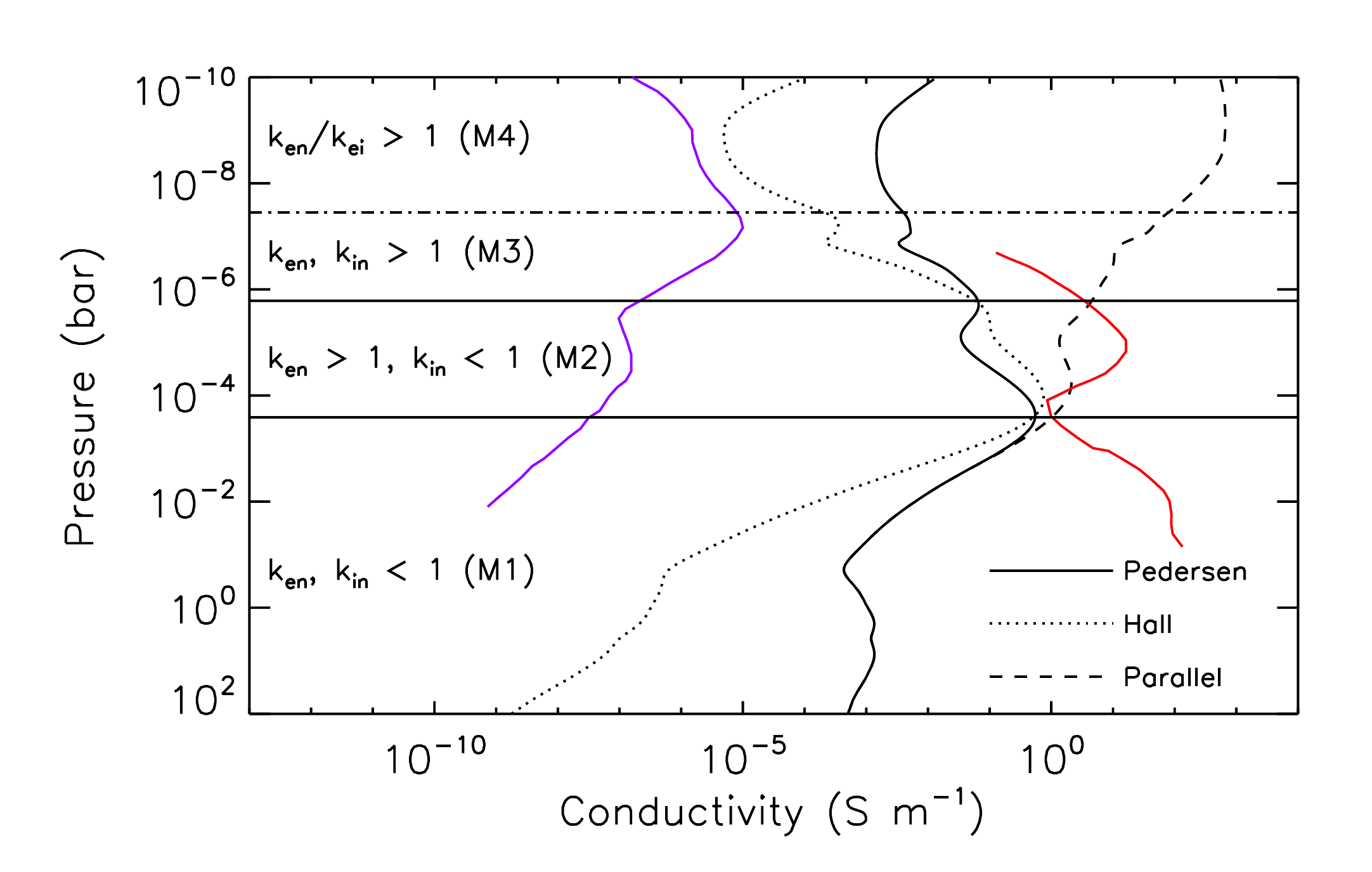}
  \caption{Conductivities in the center of mass reference frame and magnetization regions in the dayside ionosphere of HD209458b.  The red and purple lines show the Pedersen conductivities in the solar chromosphere and the ionosphere of the Earth, respectively.  The conductivities for the Earth and the Sun were taken from \citet{leake13} for illustration.\\}
  \label{fig:regimes}
\end{figure}

Figure~\ref{fig:regimes} shows the magnetization regimes and conductivities in the dayside atmosphere of HD209458b.  The atmosphere can be divided into four different regions based on the magnetizations $k_{ei}$, $k_{en}$, and $k_{in}$ of the electrons and ions.  At the highest altitudes in the M4 region $\nu_{ei} > \nu_{en}$ and the plasma behaves as if it were fully ionized.  As a general rule this is the case when the electron volume mixing ratio is $x_e \gtrsim$~10$^{-3}$, independent of the magnetic field strength.  In our model of HD209458b the transition to the fully ionized regime occurs above the 4~$\times$~10$^{-8}$ bar level in the dayside and above the 5~$\times$~10$^{-9}$ bar level in the night side.  The M4 region is not the focus of this work and hereafter we mostly concentrate on the M1, M2, and M3 regions.

\subsubsection{The M1 region}  
\label{subsc:m1}


In the lower atmosphere, below the 1 mbar level, electrons and ions are not coupled to the magnetic field lines and both $k_{en}$ and $k_{in} <<$~1.  Thus the Hall and Cowling resistivities are both small and the parallel resistivity is dominated by the $1/k_{en}$ term.  Under these circumstances conductivity is isotropic, and the induction equation~(\ref{eqn:find1}) reduces to:
\begin{equation}
\frac{\partial \mathbf{B}}{\partial t} = \nabla \times \left( \mathbf{u}_n \times \mathbf{B} - \eta_{en} \mathbf{j} \right) 
\label{eqn:perna}
\end{equation}
where $\mathbf{u}_n$ is the neutral flow velocity and the isotropic resistivity is:
\begin{equation}
\eta_{en} = \frac{m_e \nu_{en}}{e^2 n_p} = \frac{B}{e n_p} \left( \frac{1}{k_{en}} \right).
\end{equation}  
This result is also illustrated by Figure~\ref{fig:regimes} where the Pedersen conductivity is equal to the parallel conductivity in the M1 region and the Hall conductivity rapidly decreases with increasing pressure.  

We do not expect large scale currents and ion drag to be important in the M1 region.  Since both the electrons and ions are coupled to the neutral atmosphere, there is no physical mechanism to enable charge separation that is required to support currents (Section~\ref{subsc:m2m3}).  In order to see this, we summed the electron and ion momentum equations to obtain:
\begin{eqnarray}
& &n_p \left[ m_i \nu_{in} (\mathbf{u}_i - \mathbf{u}_n) + m_e \nu_{en}  (\mathbf{u}_e - \mathbf{u}_n) \right] \approx \mathbf{j} \times \mathbf{B} \nonumber \\
&\rightarrow& (\mathbf{u}_p - \mathbf{u}_n) \approx \frac{\mathbf{j} \times \mathbf{B}}{n_p m_i \nu_{in}}.
\end{eqnarray}
where $\mathbf{u}_p \approx \mathbf{u}_i$ is the plasma velocity.  If $j \approx \sigma_P u_n B$ in the M1 region, we have to an order of magnitude   
\begin{equation}
\frac{\left| u_p - u_n \right|}{u_n} \approx \frac{\sigma_P B^2}{n_p m_i \nu_{in}} = k_{en} k_{in} \approx S,
\label{eqn:veldif}
\end{equation}
which is much less than unity, implying that $u_p \rightarrow u_n$ and thus that $\mathbf{j} \rightarrow$~0.  In the M1 region, this relationship also explains the small values of $S$ that we discussed in Section~\ref{subsc:basics}.  In general, it confirms our physical intuition that ion drag is not significant in the deep atmosphere -- a result that arises partly from the proportionality of the isotropic conductivity $(1/\eta_{en})$ to electron magnetization $k_{en}$ (see also Section~\ref{sc:emdyn}).  

\subsubsection{The M2 and M3 regions}
\label{subsc:m2m3}


In the M2 region the electrons are coupled to the magnetic field ($k_{en} >$~1) while ions are coupled to the neutral atmosphere by collisions ($k_{in} <$~1).  The partial decoupling of first the electrons in the M2 region and then both the electrons and ions in the M3 region provides a natural mechanism of charge separation that is based on the interaction of the electron and ion gyromotion and collisions with the neutral wind.  In the absence of collisions electrons and ions rotate around the magnetic field lines with gyroradii that depend on their initial velocities.  Collisions with the neutral wind cause the ions and electrons to drift in opposite directions, due to their opposite sense of rotation around the magnetic field lines, and generate a current.  In the M1 region, however, collisions are so frequent that the gyromotion and the generation of significant currents are largely suppressed.   Under these circumstances both the ions and the electrons simply follow the neutral wind.        

In terms of magnetization, the M2 region is similar to the E layer in the Earth's ionosphere.  On HD209458b, however, the peak Pedersen conductivity is more than 5 orders of magnitude higher than the corresponding conductivity anywhere in the Earth's ionosphere, and actually comparable to the corresponding M2 region conductivity in the solar chromosphere.  The conductivity tensor in the M2 region is anisotropic and the Hall conductivity is higher than the Pedersen conductivity (Figure~\ref{fig:regimes}).  The parallel conductivity is generally higher than the perpendicular conductivities.  As we show in Section~\ref{sc:emdyn}, the anisotropy of the conductivity tensor and high parallel conductivity enhance ion drag and resistive heating in the upper atmosphere.  

Again in terms of magnetization, the M3 layer is equivalent to the F layer in the Earth's ionosphere, but the conductivities on HD209458b are still much higher than the conductivities in any planetary ionosphere. Both the electrons and ions are strongly coupled to the magnetic field ($k_{en} > k_{in} >$~1), and the conductivity tensor remains anisotropic.  Typically the Hall conductivity is smaller than the Pedersen conductivity and the ratio $(\sigma_{\parallel}/\sigma_P)$ is again much larger than unity.  As ions begin to separate from the neutral flow and drift in the opposite direction to the electrons, the M3 region, similarly to the M2 region, can also support strong currents.

Finally, we briefly comment on the approximations of the Ohm's law~(\ref{eqn:ohm1}) that are valid in the M2 and M3 regions.  We find that $k_{in} << k_{en}$ throughout the atmosphere.  Below the fully ionized regime it is also true that $k_{en} < k_{ei}$ and $Y_n \approx$~1.  Thus, to a fairly good approximation the parallel, Cowling and Hall resistivity reduce to:
\begin{eqnarray}
& &\eta_{\parallel} = \frac{B}{e n_p} \frac{1}{k_{en}}, \ \ \ \ \eta_C = \frac{B}{e n_p} \frac{k_{en} k_{in}}{k_{en} + k_{in}} \approx \frac{B}{e n_p} k_{in}, \nonumber \\
& &\eta_H = \frac{B}{e n_p} \frac{k_{en}-k_{in}}{k_{en} + k_{in}} \approx \frac{B}{e n_p},
\label{eqn:simple_res}
\end{eqnarray}
which are roughly valid below the 10$^{-7}$ bar level on HD209458b.  These simplifications are typical in models of solar system ionospheres \citep[e.g.,][]{schunk00} and they imply that the generalized Ohm's law in the center of mass frame reduces to the Ohm's law in the neutral frame for much of the atmosphere.

\subsection{Night side conditions}
\label{subsc:night_cond}    

\begin{figure}
  \epsscale{1.15}
  \plotone{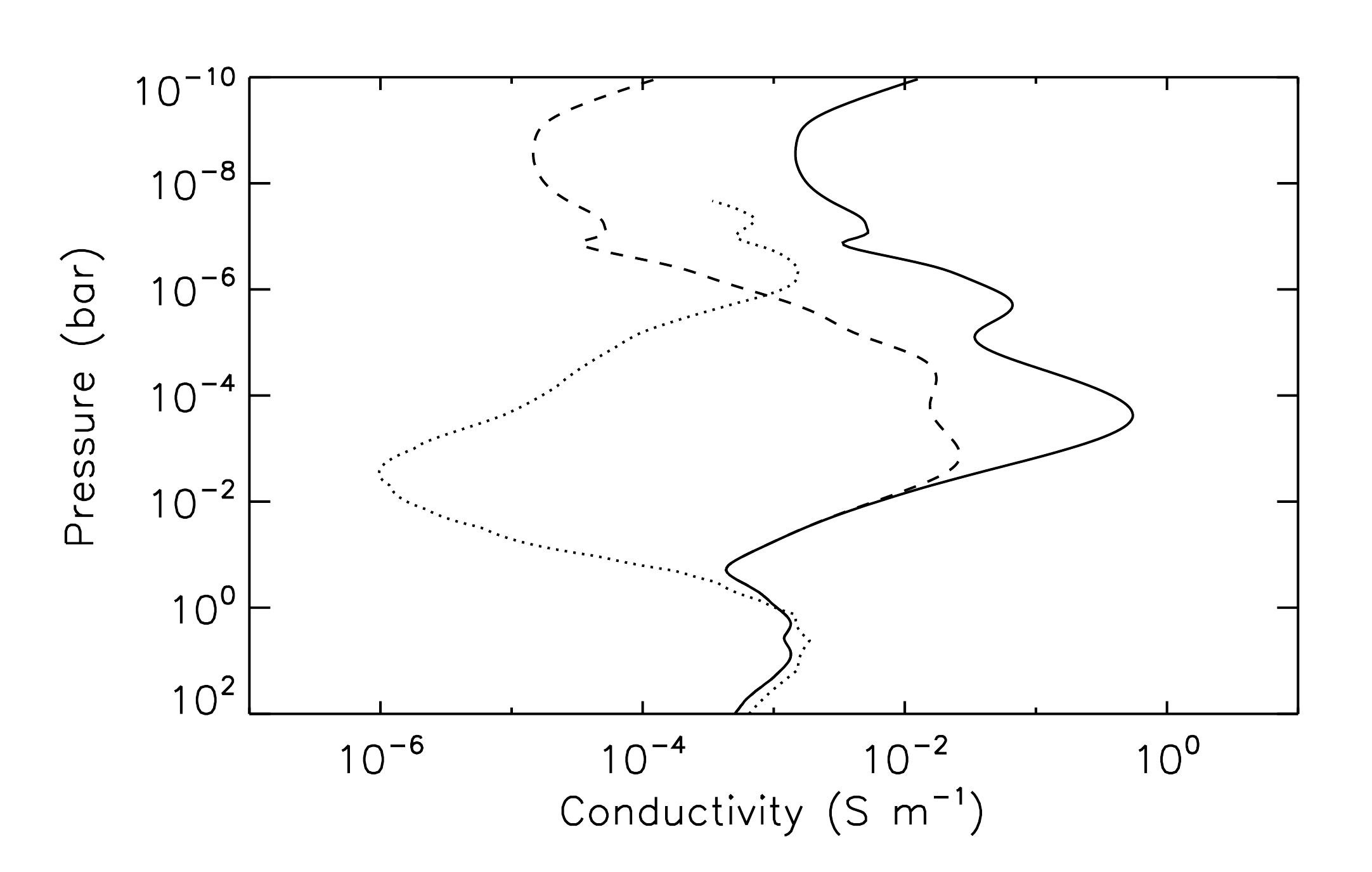}
  \caption{The difference between the Pedersen conductivity in the dayside (solid line) and night side (dotted line).  The dashed line shows the dayside conductivity for $\mu_p =$~10 $\mu_J$.}
  \label{fig:night_pedc}
\end{figure}

It is important to note that magnetization does \textit{not} depend on the electron density or conductivities -- it only depends on the strength of the magnetic field and pressure (see equation~\ref{eqn:kst1}).  For this reason, the boundaries of the M1, M2, and M3 regions are roughly the same in the night side as they are on the dayside.  The conductivities, however, are significantly different.  For example, Figure~\ref{fig:night_pedc} compares the Pedersen conductivities between the day and night sides.  Due to the differences in the electron densities, the conductivities below the 10$^{-6}$ bar level are orders of magnitude lower in the night side.  As a result, ion drag directly modifies dynamics below the 10$^{-6}$ bar level only on the dayside (see Section~\ref{subsc:magnetic2}), although the effects of strong ion drag will be felt indirectly in the night side.  

\subsection{Different magnetic field strengths}
\label{subsc:magnetic}    

\begin{figure}
  \epsscale{1.15}
  \plotone{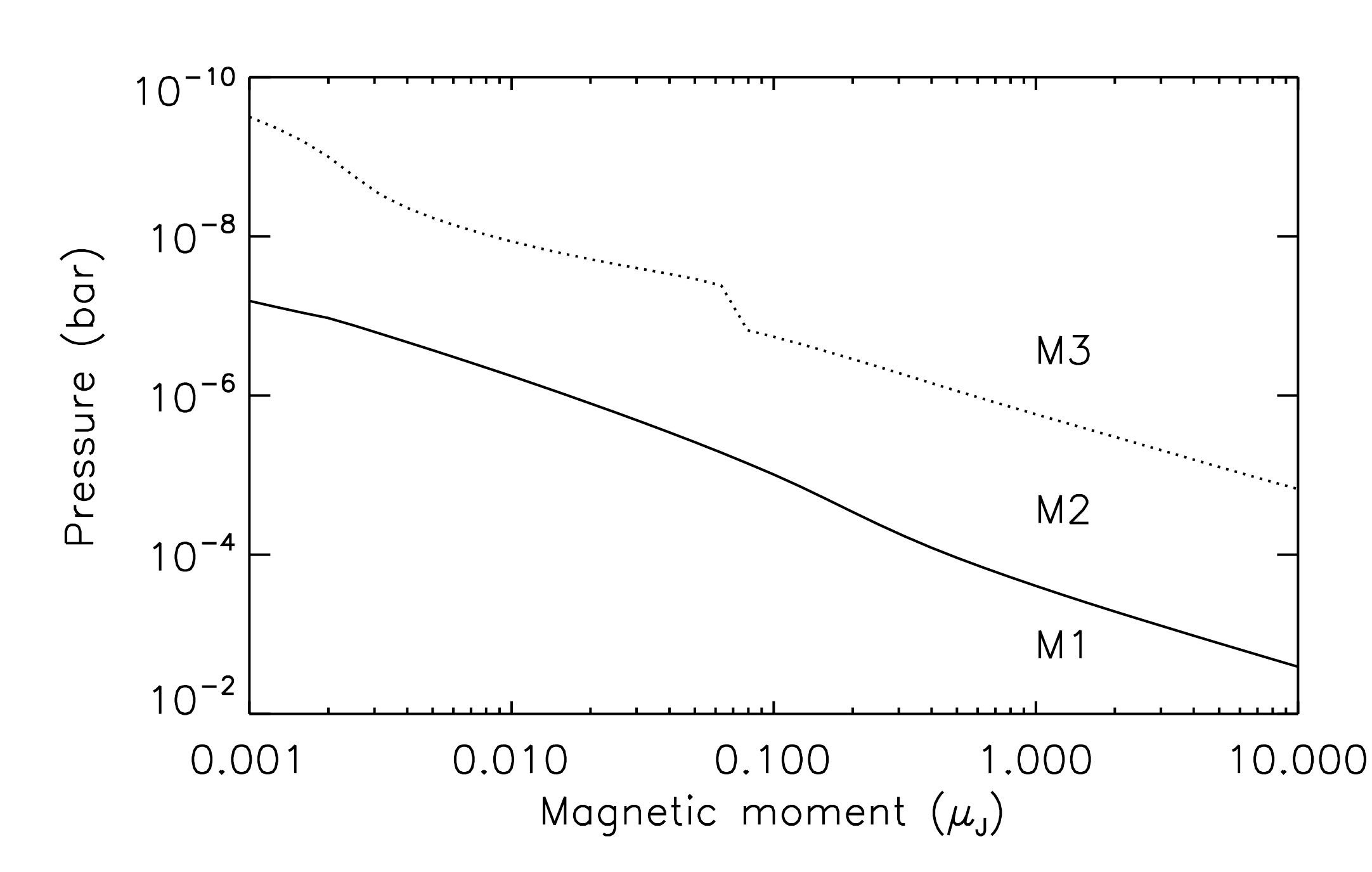}
  \caption{The lower boundary pressures of the M2 (solid) and M3 (dotted) regions as a function of the planetary magnetic dipole moment (where $\mu_J$ is the Jovian dipole moment).  The step between 0.01--0.1 $\mu_J$ in the M3 region boundary arises from the changing ion composition from metals to protons.}
  \label{fig:bfield}
\end{figure}

The boundaries of the M1, M2, and M3 regions obviously depend on the planetary magnetic field strength.  In order to illustrate this, we calculated magnetization for different dipole moments ranging from 0.01 $\mu_J$ to 10 $\mu_J$ (equatorial surface fields $B_0$ from 1.8~$\times$~10$^{-6}$ T to 1.8~$\times$~10$^{-3}$ T).  Figure~\ref{fig:bfield} shows the lower boundaries of the M2 and M3 regions as a function of dipole moment.  It implies that for a reasonable range of dipole moments the M2 and M3 regions are always located above the 0.01 bar level.  Given that scaling relations such as equation~(\ref{eqn:veldif}) are valid regardless of the electron density, this is the pressure range where ion drag is potentially significant.  For very low magnetic moments of $\mu_p \lesssim$~0.01 $\mu_J$ \citep[e.g.,][]{griesmeier04}, on the other hand, the M2 region is located entirely in the thermosphere and the M3 region merges into the M4 region (i.e., the `fully ionized' regime).           

Higher magnetic fields expose more of the atmosphere to ion drag, but the general effect of the surface field strength on the conductivities deserves further scrutiny.  Clearly $\sigma_{\parallel}$ does not depend on $B_0$.  Pedersen conductivity, on the other hand, is always equal to $\sigma_{\parallel}$ in the M1 region and less than $\sigma_{\parallel}$ in the M2 region.  Thus $\sigma_P$ \textit{decreases} with increasing $B_0$ in the M2 and M3 regions (Figure~\ref{fig:night_pedc}).  Similarly, the Hall conductivity $\sigma_H$ decreases with increasing $B_0$ in the M2 and M3 regions while it increases with $B_0$ in the M1 region.  The overall effect of increasing $B_0$ is thus to enhance the degree of anisotropy in the conductivity tensor.  We note, though, that increasing $B_0$ leads to stronger ion drag despite the reduced conductivities (and perpendicular currents) because ion drag also depends directly on $B_0$, and this compensates for the reduced conductivities (Section~\ref{subsc:magnetic2}).
 
\section{Currents and ion drag}
\label{sc:emdyn}    

In this section we use a steady state Ohm's law to confirm that currents and ion drag can significantly affect the dynamics and energy balance in the M2 and M3 regions.  Our main focus is on the anisotropic conductivity that has not been considered in previous studies of exoplanet atmospheres.  Under the general assumption that $\nabla \cdot \mathbf{j} =$~0, equation~(\ref{eqn:ohm2}) can be written as:
\begin{equation}
\nabla \cdot (\overline{\sigma} \cdot \mathbf{E}) = -\nabla \cdot [ \overline{\sigma} \cdot (\mathbf{u}_n \times \mathbf{B} ) ].
\label{eqn:estatic1}
\end{equation}  
This equation can be solved self-consistently during each time step within a GCM to obtain a realistic description of ion drag.  To the best of our knowledge, GCMs that include ion drag in this manner have not been developed for any other planet than the Earth \citep[e.g.,][]{richmond00}.  The coupling of electrodynamics to realistic circulation models is indeed a complex undertaking that we do not pursue here.  Instead we concentrate on a few simple examples to demonstrate the qualitative effect of electrodynamics on planetary atmospheres.  In all cases below we work in the reference frame of the neutral atmosphere.       

\subsection{Mid-latitude jet}
\label{subsc:mid_lat_jet}

\begin{figure}
  \epsscale{1.15}
  \plotone{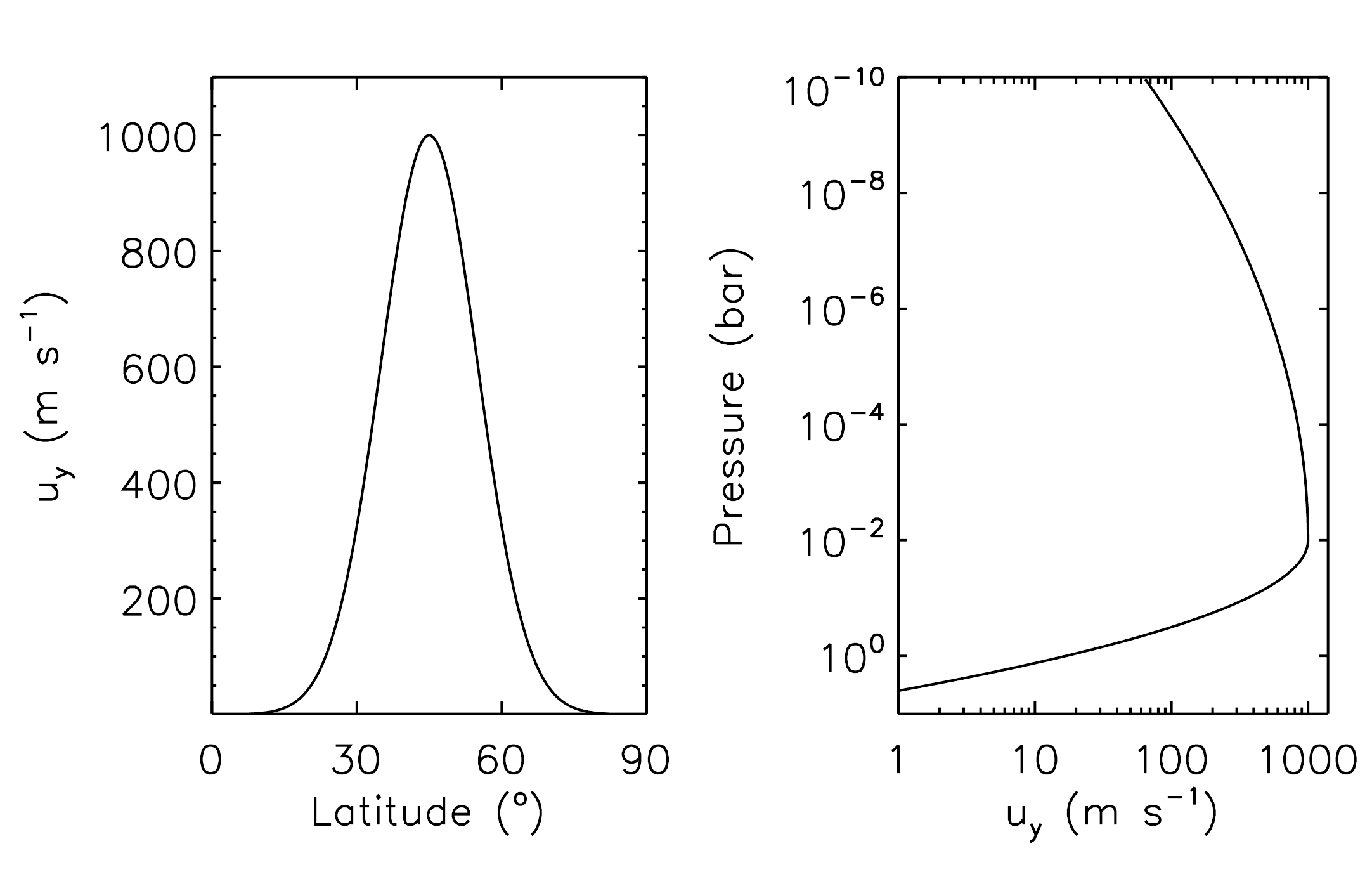}
  \caption{The simplified mid-latitude zonal jet profile as a function of latitude (left panel) and pressure (right panel) that we use for electrodynamics demonstrations.}
  \label{fig:box_wind}
\end{figure}

Mid-latitude zonal jets arise naturally on rotating planets from geostrophic balance.  The best known example is the jet stream on the Earth \citep[e.g.,][]{salby96}.  We consider a highly idealized eastward jet in the northern hemisphere with a peak zonal wind speed of $u_{y0} =$~1 km~s$^{-1}$ that is constant in longitude and follows a Gaussian profile in latitude and a pressure profile shown in Figure~\ref{fig:box_wind}.  Although this wind profile is not predicted by any exoplanet circulation model, the pressure dependency of the jet is motivated by such models and the wind speed is designed to reach maximum at 0.01 bar, reduce to zero at 10 bar \citep[e.g.,][]{showman09} and decrease to about 500 m~s$^{-1}$ near the 10$^{-6}$ bar level \citep{koskinen10}.  We note that these properties are typical of an equatorial jet on EGPs, but here we adapt them to a mid-latitude jet for illustration purposes.  We consider an equatorial jet with the same pressure and latitude dependency in Section~\ref{subsc:eq_jet}.   

\begin{figure}
  \epsscale{1.15}
  \plotone{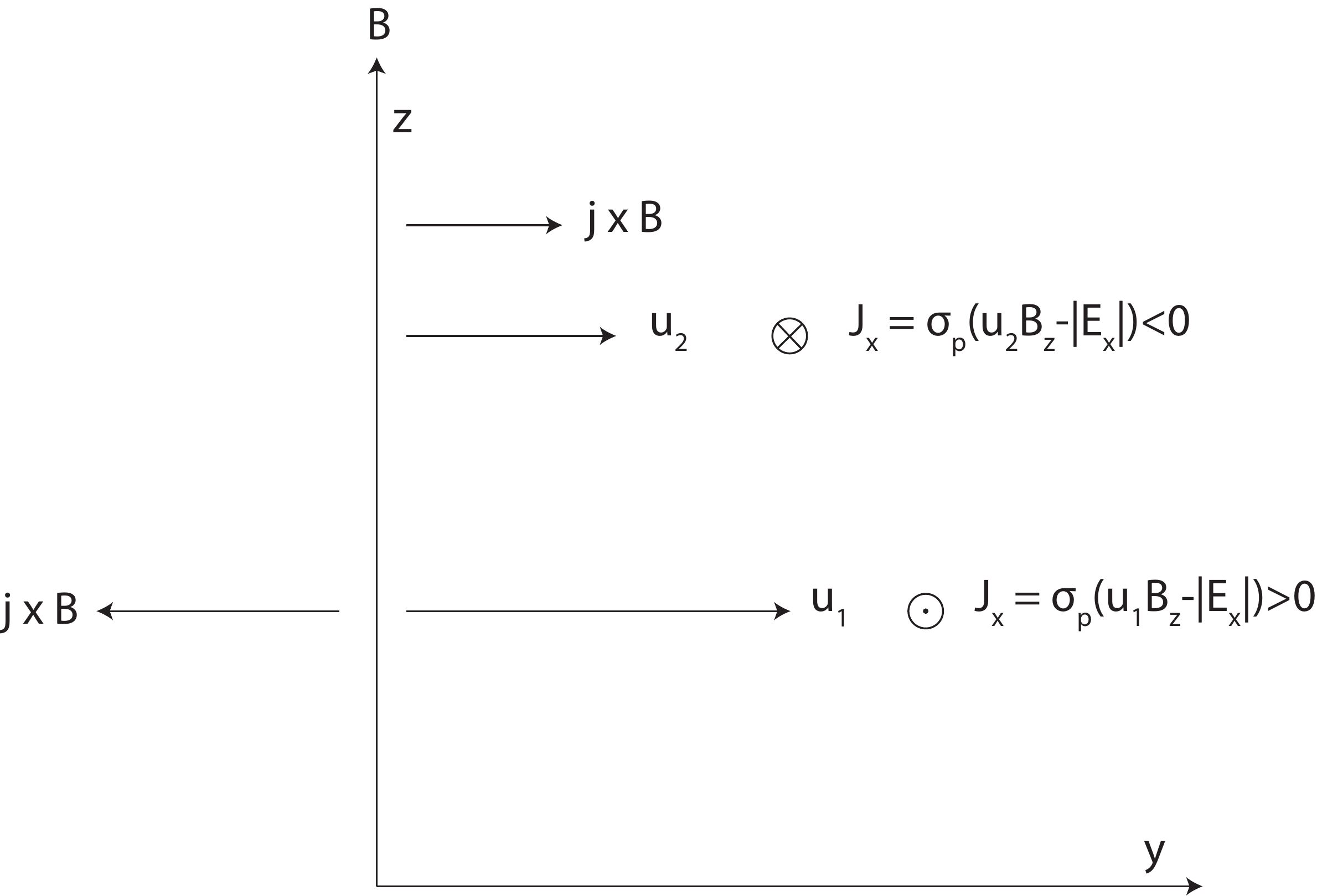}
  \caption{Illustration of the meridional currents and electric fields associated with an eastward mid-latitude jet in the northern hemisphere (Section~\ref{subsc:mid_lat_jet}).  In this example we consider an upward magnetic field line (with the $x$ axis pointing south).  When $\sigma_{\parallel}/\sigma_P >>$~1, the electric field $E_x$ is constant along the magnetic field line.  When $B_z = B_0$ is constant, $u_1 B_0 > | E_x | > u_2 B_0$ and the current densities based on $u_1$ and $u_2$ are positive and negative, respectively.  Thus ion drag accelerates $u_2$ and decelerates $u_1$ until the wind speeds are equal.  Note that the subscripts 1 and 2 here indicate different wind speeds and not the M1 and M2 regions.}
  \label{fig:box_geometry}
\end{figure}

At mid-latitudes the magnetic dipole field lines are almost vertical and the magnetic field coordinates can roughly be reduced to Cartesian coordinates\footnote{In this system the $z$ axis lies in the direction of the magnetic field, which defines the directions of the $x$ and $y$ axes.}.  The components of the current density are:
\begin{eqnarray}
j_x &=& \sigma_P (E_x + u_y B) - \sigma_H (E_y - u_x B)  \nonumber \\
j_y &=& \sigma_H (E_x + u_y B) + \sigma_P (E_y - u_x B) \nonumber \\
j_z &=& \sigma_{\parallel} E_z 
\label{eqn:currents}
\end{eqnarray}  
where $x$ is the meridional dimension and $y$ is the zonal dimension (Figure~\ref{fig:box_geometry}).  We assume that the meridional wind ($u_x$) is negligible, the zonal wind ($u_y$) is constant in longitude, and the conductivities only change with altitude.  In solving equations~(\ref{eqn:currents}) we also assume that the magnetic field strength $B_0 =$~1.8~$\times$~10$^{-4}$ T (1.8 G) is constant with latitude and altitude.  Thus $\nabla \cdot \mathbf{j} =$~0 yields:
\begin{eqnarray}
& &\sigma_P \frac{\partial}{\partial x} \left( E_x + u_y B_0 \right) + \sigma_P \frac{\partial E_y}{\partial y} \nonumber \\
& &- \sigma_H \left( \frac{\partial E_y}{\partial x} - \frac{\partial E_x}{\partial y} \right) + \frac{\partial}{\partial z} \left( \sigma_{\parallel} E_z \right) = 0.
\label{eqn:estatic1b}
\end{eqnarray}
Despite the fact that $S \rightarrow$~1 in the upper atmosphere, we use the static approximation with $\nabla \times \mathbf{E} =$~0 and write:    
\begin{equation}
\frac{\partial^2 \Phi}{\partial x^2} + \frac{1}{\sigma_P} \frac{\partial}{\partial z} \left( \sigma_{\parallel} \frac{\partial \Phi}{\partial z} \right) = \frac{\partial}{\partial x} (u_y B_0) = S(x,z)
\label{eqn:estatic2}
\end{equation}  
where $\Phi$ is the electrostatic potential, $\mathbf{E} = -\nabla \Phi$ and $S(x,z)$ is the source term.  We note that the symmetry of this example means that $\Phi$ is constant with longitude and thus $E_y =$~0.

\begin{figure}
  \epsscale{1.15}
  \plotone{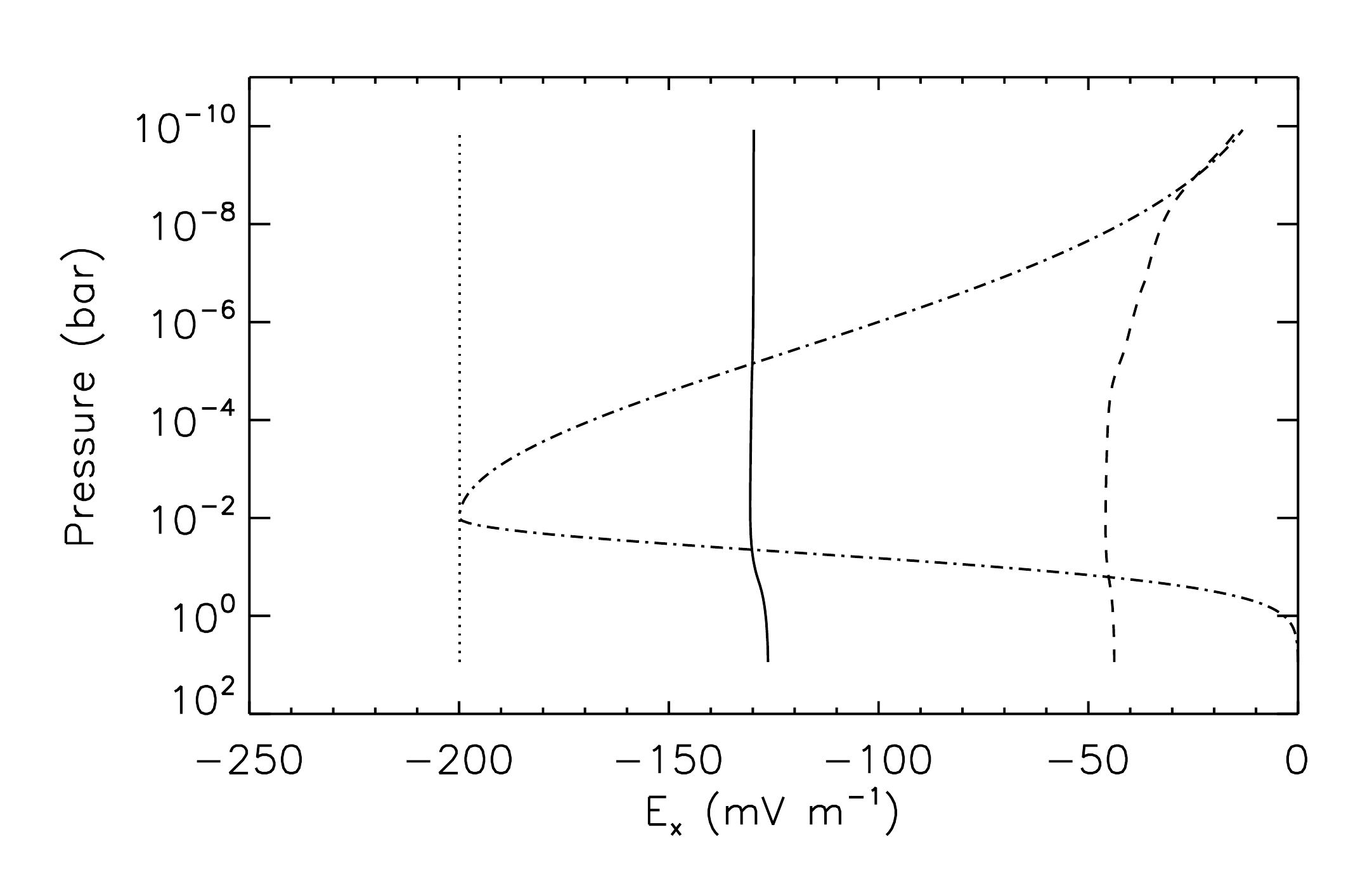}
  \caption{Polarization electric fields based on three different cases of the mid-latitude jet and conductivity profiles (see Section~\ref{subsc:mid_lat_jet}).  The different solutions are based on constant conductivity and constant $u_y$ with altitude (dotted line), full wind profile (Figure~\ref{fig:box_wind}) and variable conductivity with $\sigma_{\parallel} = \sigma_P$ (dashed line), and full wind profile and anisotropic conductivity (solid line).  The field $E_x = -u_y B_z$ based on the full wind profile is also shown (dash-dotted line).}
  \label{fig:efields1}
\end{figure}

We solve equation~(\ref{eqn:estatic2}) numerically for $\Phi$ at pressures ranging from 10 bar to 10$^{-10}$ bar with zero current boundary conditions.  Such boundary conditions are not appropriate in all circumstances.  The aim of this section, however, is to provide useful qualitative insight to atmospheric electrodynamics and these boundary conditions are sufficient for this purpose.  As a result, Figure~\ref{fig:efields1} shows the polarization electric field $E_x = -\partial \Phi/\partial x$ at $\lambda =$~45$^{\circ}$ for three cases that highlight different aspects of electrodynamics.  In the first case we have assumed that the zonal jet is constant in pressure, and set $\sigma_{\parallel} = \sigma_P =$~1 S~m$^{-1}$ everywhere (dotted line).  This case shows that currents vanish and there is no ion drag when both the wind speed $u_y$ and $B_z$ are constant with $z$ because under these circumstances equation~(\ref{eqn:estatic2}) yields $E_x = -u_y B_0$ everywhere.  This leads to the important conclusion that ($-\mathbf{u}_n \times \mathbf{B}$) must change along the magnetic field lines for there to be any appreciable ion drag.

The second case demonstrates the general result that ion drag attempts to eliminate any variation of $(-u_y B_z)$ with altitude (i.e., along the magnetic field lines).  In this case we used the full wind profile (Figure~\ref{fig:box_wind}) and the dayside conductivities (Figure~\ref{fig:regimes}).  The resulting polarization field $E_x$, which is nearly constant with altitude, is shown by the solid line in Figure~\ref{fig:efields1}.  In this case ($-u_y B_0$) changes with altitude and thus the potential $\Phi$ must also change with altitude.  Due to the diffusive effect of the left hand side in equation~(\ref{eqn:estatic2}) the polarization field $E_x \ne -u_y B_0$.  As a result, the meridional current at $\lambda =$~45$^{\circ}$ is positive between 8~$\times$~10$^{-6}$ bar and 0.04 bar and negative at other pressure levels.  Thus ion drag decelerates the zonal wind around the peak altitude of the wind profile and accelerates it at other altitudes.  If $B_z$ is constant in altitude, $u_y$ will tend to a constant value with altitude at each latitude.  If $B_z$ is not constant, the vertical wind profiles tend to the structure of the magnetic field based on equation~(\ref{eqn:estatic2}).    

We note that although $E_y =$~0, there is a zonal current that arises from the Hall effect with $j_y = \sigma_H (E_x + u_y B_0)$ (equation~\ref{eqn:currents}), which is constant with longitude.  Given that $\sigma_H > \sigma_P$ in the M2 region, this current is often comparable to or stronger than the meridional current.  The Hall current flows perpendicular to both the magnetic field and the electric field and thus it does not contribute to resistive heating of the atmosphere (Section~\ref{subsc:joule}).  It does, however, lead to ion drag on the meridional winds.  Since we consider the meridional wind negligible and generally do not concern ourselves with the meridional momentum balance in this example, we do not discuss the Hall currents any further here.

The third example (dashed line in Figure~\ref{fig:efields1}) demonstrates the error of ignoring the anisotropy of the conductivity tensor.  This case is otherwise identical to the second example above but we set $\sigma_P = \sigma_{\parallel}$.  The resulting electric field is up to 10 times lower in the thermosphere and 2--3 times lower throughout the atmosphere.  Interestingly the effect of anisotropic conductivity is not limited to the M2 and M3 regions because the lower atmosphere has to adapt to the electric fields in the upper atmosphere in regions where the magnetic field lines traverse through the whole atmosphere.  This effect is potentially large and it cannot be ignored, even when it complicates the electrodynamics significantly.  For the remainder of this section we concentrate on the second example above.

\begin{figure}
  \epsscale{1.15}
  \plotone{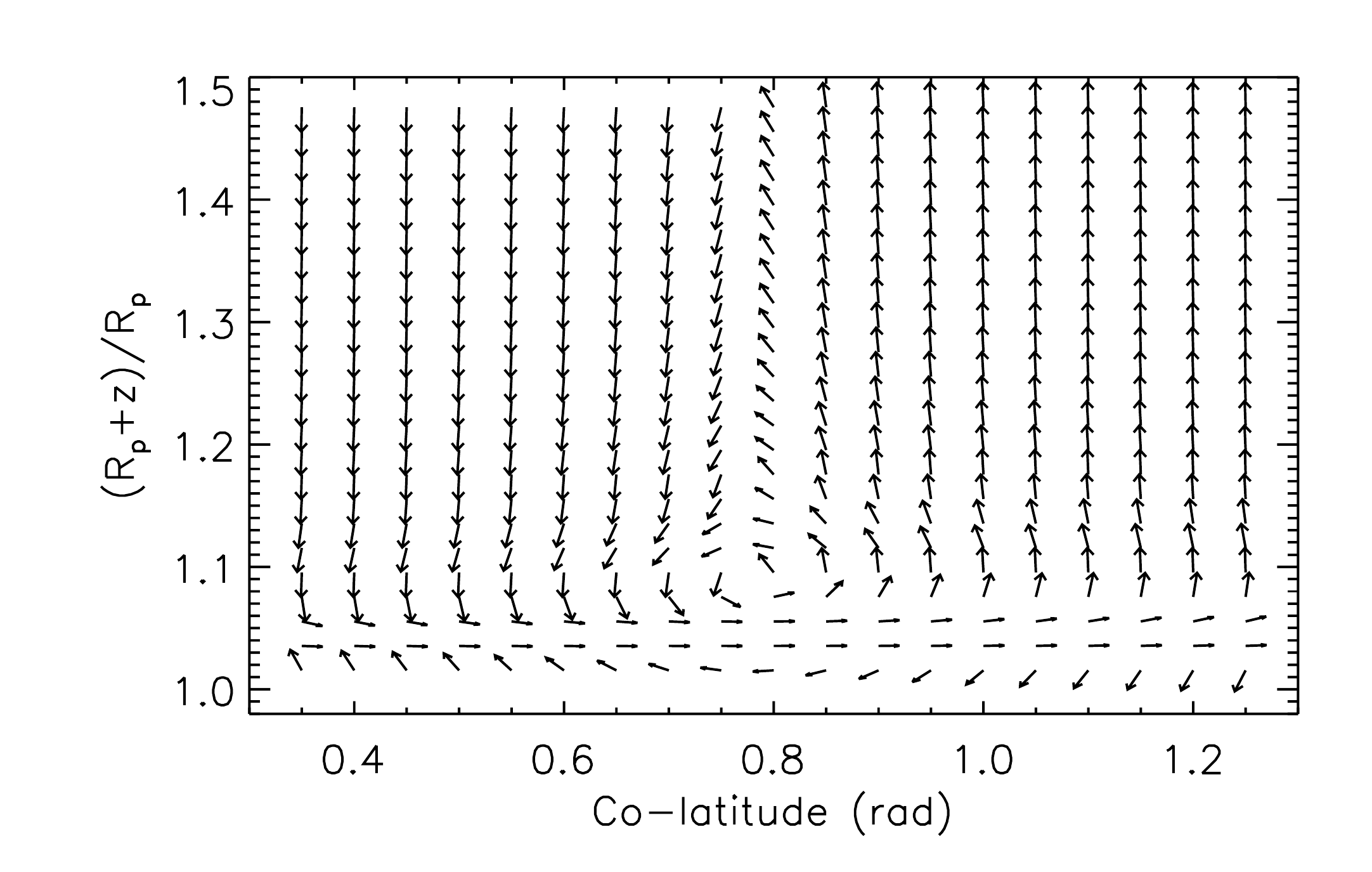}
  \caption{Current loop established by the mid-latitude jet.  Note that the arrows in the left panel depict unit vectors and the magnitude of the current changes significantly with location (e.g., Figure~\ref{fig:jx1b}).}
  \label{fig:jx1a}
\end{figure}      

\begin{figure}
  \epsscale{1.15}
  \plotone{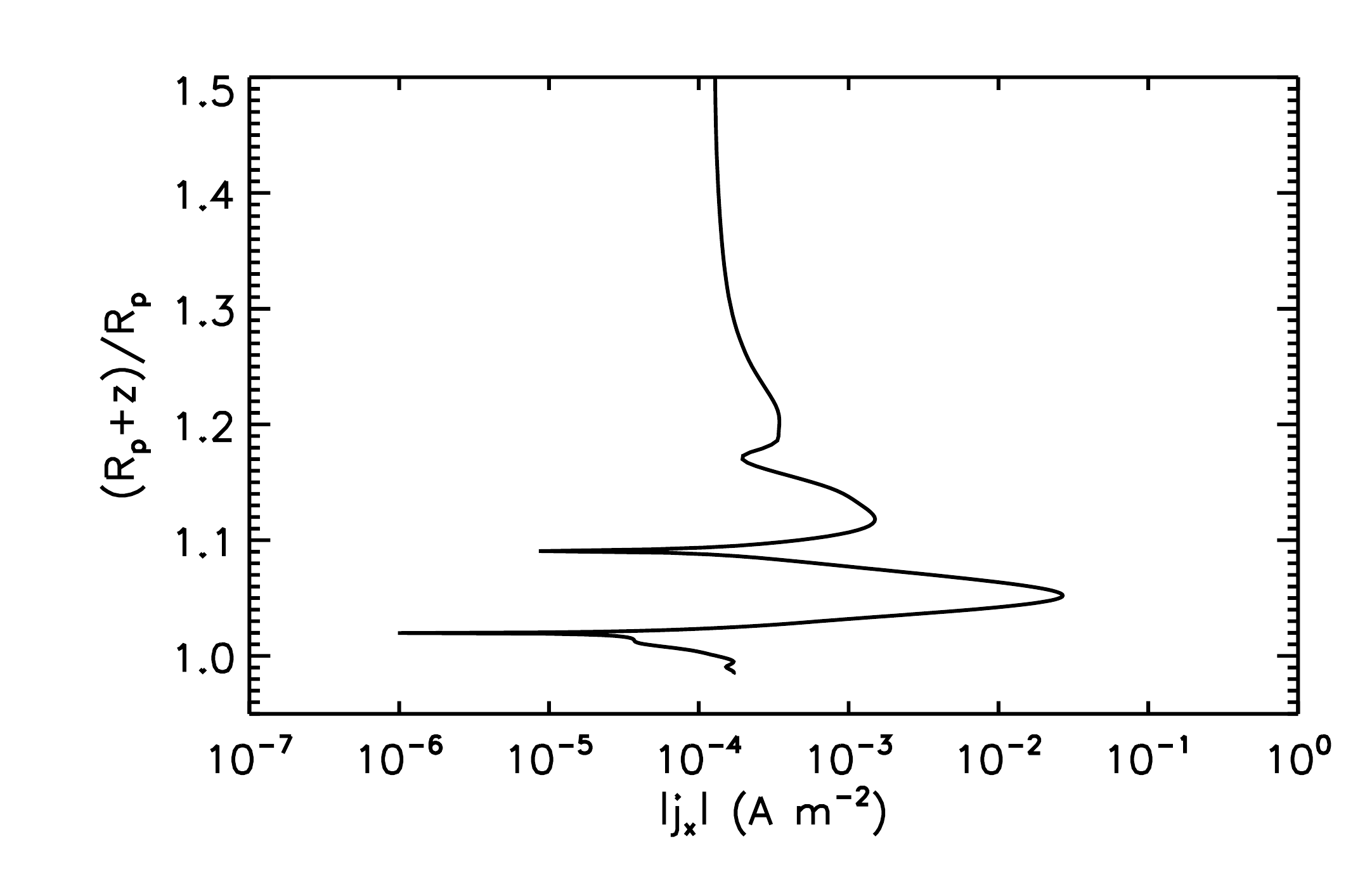}
  \caption{Meridional current density $| j_x |$ at $\lambda =$~45$^{\circ}$ based on the mid-latitude jet.  The direction of the current is shown in Figure~\ref{fig:jx1a}.} 
  \label{fig:jx1b}
\end{figure}

As we stated above, ion drag attempts to make $u_y$ constant with altitude at each latitude within the jet.  The current loop responsible for this is shown by Figure~\ref{fig:jx1a} while the corresponding current density at $\lambda =$~45$^{\circ}$ is shown by Figure~\ref{fig:jx1b}.  With upward vertical field lines in the northern hemisphere, the current flows anti-clockwise in the upper atmosphere and clockwise in the lower atmosphere.  The current density at $\lambda =$~45$^{\circ}$ varies from about 0.1 mA~m$^{-2}$ to a maximum of 30 mA~m$^{-2}$.  In terms of latitude, the current is strongest near the peak of the wind profile and decreases away from the peak (not shown).  Thus deceleration around the 0.01 bar level and acceleration in the upper and lower atmospheres is strongest at $\lambda =$~45$^{\circ}$.  As a result, we expect a flattened version of the original zonal jet from the 0.01 bar level to appear over a wide range of altitudes if ion drag is sufficiently strong to affect the momentum balance.     

As we argued before, ion drag can dominate the dynamics on close-in EGPs in the M2 and M3 regions.  To see this we used the ratio of the ion drag term $\mathbf{j} \times \mathbf{B}/\rho$ in the neutral momentum equation to the pressure gradient to define an ion drag length:
\begin{equation}
l_d = \frac{\rho c_s^2}{j_x B}
\label{eqn:ld1}
\end{equation}
where $c_s^2 = k T/m$.  This is the \textit{shortest} length scale at which acceleration by ion drag overtakes acceleration by the pressure gradients or advection.  The definition can be generalized by assuming that $j_x \approx \sigma_P u_n B$ \citep[e.g.,][]{menou12}, in which case:
\begin{equation}
l_d = \frac{1}{\mu_0 u_n \sigma_P} \left( \frac{c_s}{V_{An}} \right)^2.
\label{eqn:ld2}
\end{equation} 

\begin{figure}
  \epsscale{1.15}
  \plotone{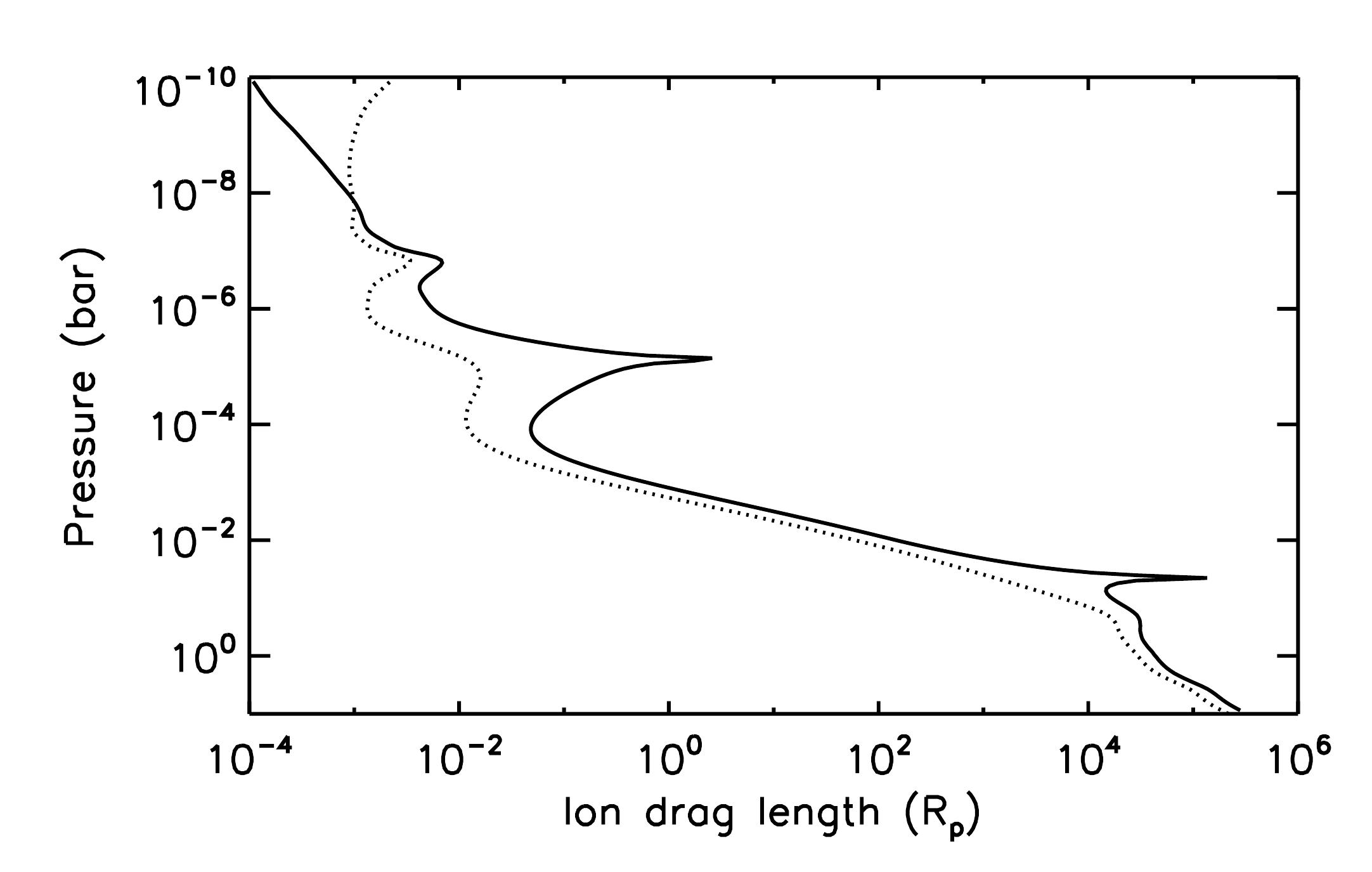}
  \caption{The ion drag length $l_d$ at $\lambda =$~45$^{\circ}$ based on the current associated with the mid-latitude jet (solid line) and the assumption that $j_x = \sigma_P u_n B$ (dotted line).  Note that the sharp peaks of the solid line are regions where the current changes sign.\\}
  \label{fig:box_mom}
\end{figure}

Figure~\ref{fig:box_mom} shows $l_d$ based on equations~(\ref{eqn:ld1}) and (\ref{eqn:ld2}).  As expected, it indicates that $l_d << R_p$ above the 10$^{-3}$ bar level.  In this region the neutral momentum balance is dominated by pressure gradients (that arise from variations in stellar heating) and ion drag.  The altitude peak of the jet, however, is at the 0.01 bar level where $l_d \approx$~150 $R_p$, suggesting that ion drag forces are less than 10$^{-2}$ of the pressure gradient forces.  This means that circulation driven by stellar irradiation in the lower atmosphere can strongly interfere with circulation in the upper atmosphere while being little affected itself.  The generalized version of $l_d$ tends to overestimate the importance of ion drag, although there can also be regions where it underestimates ion drag.  Nevertheless, in this case the agreement between the two $l_d$ is quite good.  We note that Figure~\ref{fig:box_mom} also confirms the conclusion in Section~\ref{subsc:m1} that ion drag is not important in the M1 region where $l_d >> R_p$.       

Finally, we discuss an important simplification that arises from the near-constancy of $E_x$ along the magnetic field lines when $\sigma_{\parallel} >> \sigma_P$.  Under such conditions equation~(\ref{eqn:estatic1b}) can be integrated along the field lines and solved directly for the polarization field \citep[e.g.,][]{richmond95}:
\begin{equation}
E_x = - \frac{\int_z \sigma_P u_y B \  \text{d} z}{\int_z \sigma_P \text{d} z}.
\label{eqn:estatic3}
\end{equation}   
This formula yields an electric field of $E_x =$~-130 mV~m$^{-1}$ at $\lambda =$~45$^{\circ}$, which is the same as the actual electric field in Figure~\ref{fig:efields1}.  This justifies our use of equation~(\ref{eqn:estatic3}) below in Section~\ref{subsc:eq_jet} where we consider the currents associated with an equatorial zonal jet.

\subsection{Equatorial jet}
\label{subsc:eq_jet}    

\begin{figure}
  \epsscale{1.15}
  \plotone{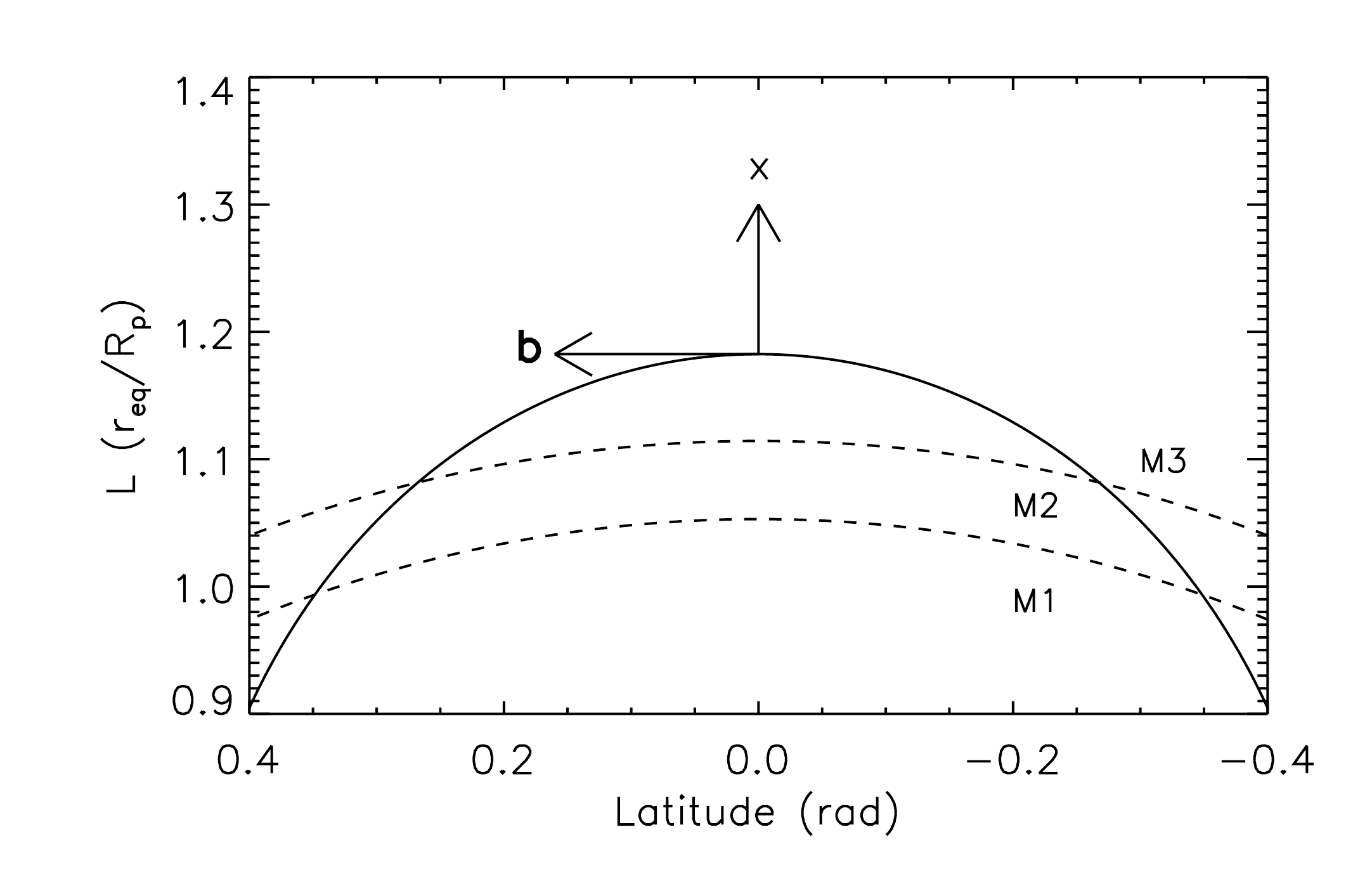}
  \caption{The solid line shows the dipole field line ($r/R_p = L \cos^2 \lambda$) with the L ($=r_{\text{eq}}/R_p$) value corresponding to the 10$^{-7}$ bar level.  North is in the direction of the magnetic field vector $\mathbf{b}$ at the equator where the magnetic $x$ axis points radially outward.  At higher latitudes the magnetic $x$ axis has both radial and meridional components.  The boundaries of the magnetization regions, which are assumed to be spherically symmetric, are shown by the dashed lines.  The magnetic $y$ axis points (eastward) into the page.}
  \label{fig:geom1}
\end{figure}

In this section we consider an equatorial zonal jet, which is a more realistic example on close-in EGPs and can be used to highlight the effects of a more complicated magnetic field geometry (Figure~\ref{fig:geom1}).  We assume the same pressure and latitude dependency as before (Figure~\ref{fig:box_wind}), but with the peak at the equator.  We reduce the problem to two dimensions and assume again that $u_y$ is constant with longitude while ignoring vertical and meridional winds.  We write equation~(\ref{eqn:estatic1b}) in a right-handed coordinate system defined by the direction of the magnetic field lines ($\mathbf{b}$) as:
\begin{equation}
\frac{\partial}{\partial x} ( \sigma_P E_x + \sigma_P u_y B) = - \frac{\partial}{\partial b} (\sigma_{\parallel} E_b).
\label{eqn:estatic4}
\end{equation}
where the $x$ direction is always perpendicular to the local magnetic field line and $y$ is the zonal direction.  The path length element $\text{d} b$ along the magnetic dipole field lines is:
\begin{equation}
\text{d} b = r_{\text{eq}} \cos (\lambda) \sqrt{ 1+3 \sin^2(\lambda) } \ \text{d} \lambda
\end{equation}
where $r_{eq}$ is the radial distance to the field line at the equator and $\lambda$ is latitude.  \textit{It is important to note that, in contrast to Section~\ref{subsc:mid_lat_jet}, we assume a south to north orientation for the magnetic field in this example}.

If the field-aligned currents vanish at the footpoints of the magnetic field lines, we can use equation~(\ref{eqn:estatic4}) to obtain the same simplification for $E_x$ as in equation~(\ref{eqn:estatic3}), with $\text{d} z$ replaced by $\text{d} b$.  The boundary conditions in this case are somewhat more defensible than the zero current boundary conditions in the mid-latitude case.  The use of equation~(\ref{eqn:estatic3}), however, leads to some limitations on our results.  Section~\ref{subsc:mid_lat_jet} indicates that this equation is roughly valid along magnetic field lines that pass to the region where $\sigma_{\parallel} > \sigma_P$.  At mid-latitudes this is the case for all field lines but at the equator only the field lines with L values ($L = r_{\text{eq}}/R_p$) in the M2 or M3 regions qualify (e.g., Figure~\ref{fig:geom1}).  Thus we only show results above the 10$^{-3}$ bar level in this section.

\begin{figure}
  \epsscale{1.15}
  \plotone{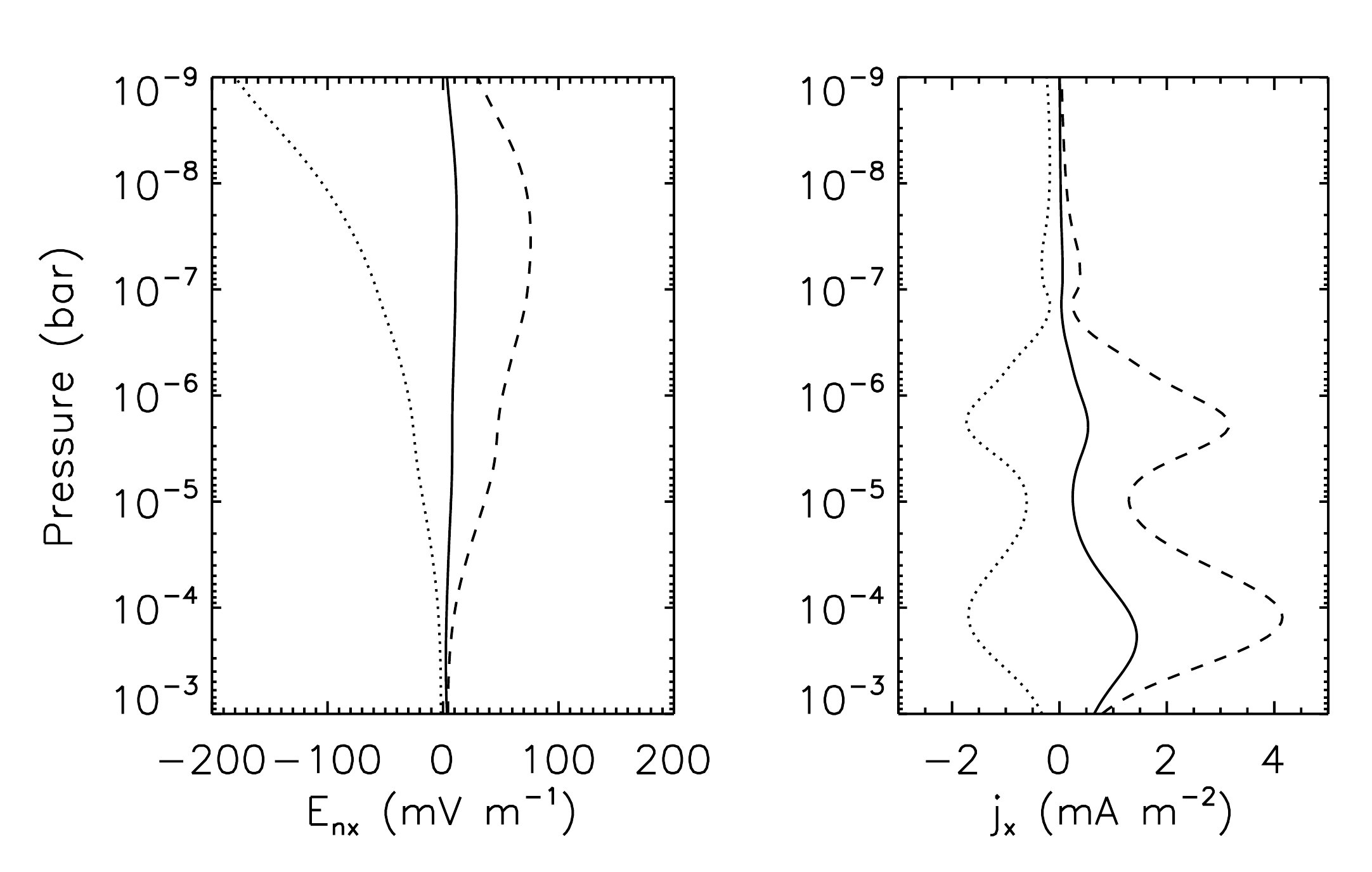}
  \caption{The net electric field perpendicular to the magnetic field lines in the neutral frame (left panel) and current density (right panel) at the equator (see text).  Three examples are shown: equatorial zonal jet where the wind varies with latitude and pressure (solid line), an equatorial jet that is constant with altitude (dashed line) and constant zonal wind everywhere (dotted line).}
  \label{fig:efields2}
\end{figure}

\begin{figure}
  \epsscale{1.15}
  \plotone{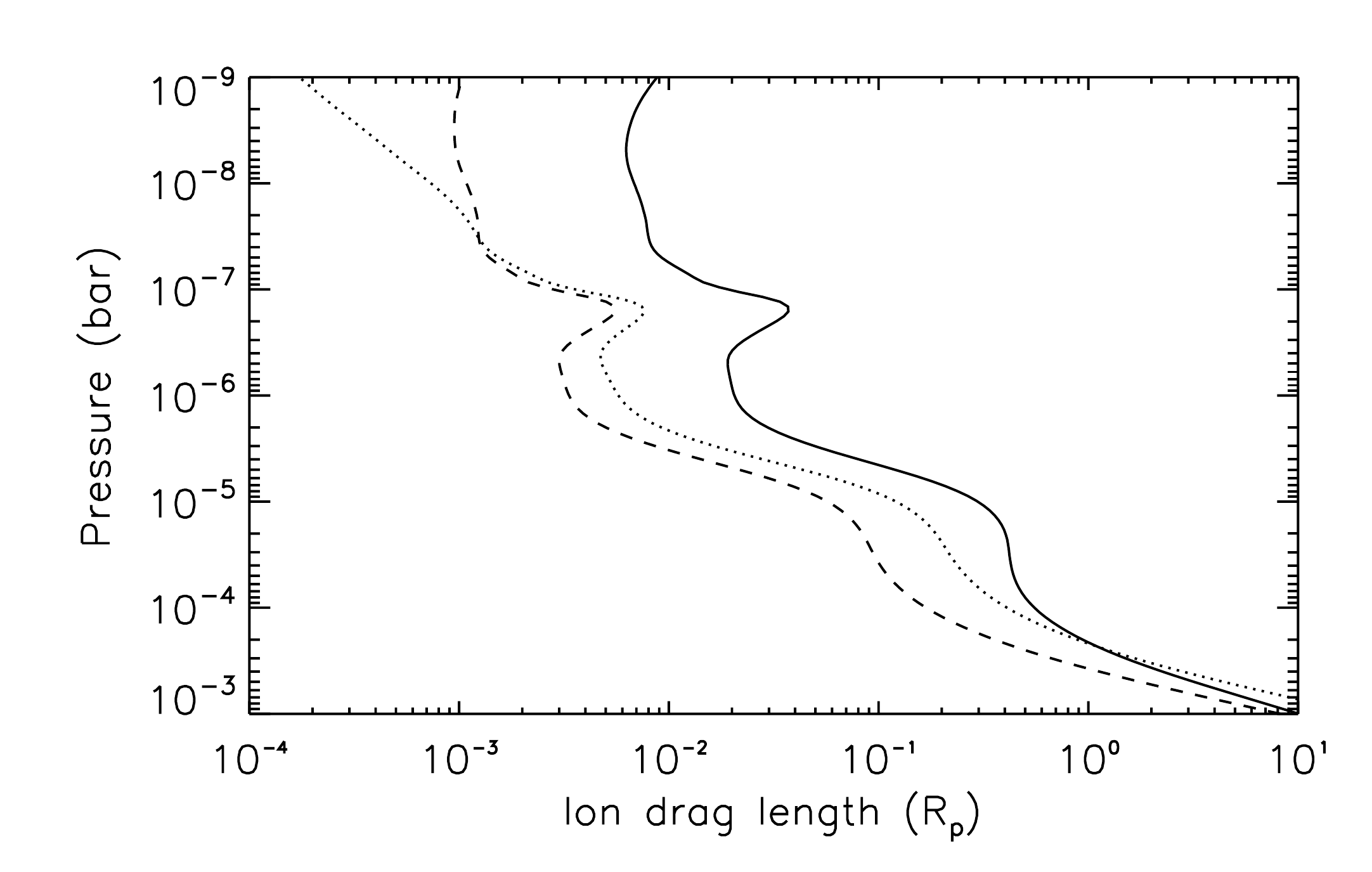}
  \caption{Ion drag lengths $l_d$ based on the three equatorial current density profiles in Figure~\ref{fig:efields2}.}
  \label{fig:eq_mom}
\end{figure}

Figure~\ref{fig:efields2} shows the net electric field perpendicular to $\mathbf{b}$ in the neutral rest frame ($E_{nx} = E_x + u_y B$) at the equator with the associated current density $j_x$ for three different cases that again highlight different aspects of electrodynamics.  The corresponding values of $l_d$ are shown in Figure~\ref{fig:eq_mom}.  All of these examples are based on the dayside conductivity profiles from Figure~\ref{fig:regimes}.  In the first case the zonal wind $u_y =$~1 km~s$^{-1}$ is constant everywhere (dotted line).  In the second case the zonal wind speed varies with latitude but remains constant in pressure (dashed line).  Finally, in the third case we use the full wind profile (Figure~\ref{fig:box_wind} with the peak at the equator) that varies with both latitude and pressure (solid line).  Similarly to Section~\ref{subsc:mid_lat_jet}, all of these cases produce zonal Hall currents in the M2 region.  The Hall currents lead to ion drag on the meridional and vertical winds that we ignore in this section.

The first case demonstrates that spatial variations in the dipole magnetic field alone can drive significant ion drag in the upper atmosphere.  For constant $u_y$, the polarization field $E_x$ along the magnetic field lines is given by:
\begin{equation}
E_x (L)  = -u_y \frac{\int_b \sigma_P B \text{d} b}{\int_b \sigma_P \text{d} b} = -u_y \overline{B}_L
\end{equation}  
where $\overline{B}_L$ is the conductivity-weighted mean magnetic field strength along the field line with $L = r_{\text{eq}}/R_p$.  The dipole magnetic field strength along the field lines with a given L value is:
\begin{equation}
B(\lambda,L) = \frac{B_{\text{eq}} (R_p) \sqrt{1 + 3 \sin^2 \lambda}}{ L^3 \cos^6 \lambda}.
\end{equation}
We note that $\overline{B}_L$ is heavily weighted towards values around the peak of $\sigma_P$ near $p =$~2~$\times$~10$^{-4}$ bar (Figure~\ref{fig:regimes}).  For example, field lines with $L = 1.124$ ($p_{\text{eq}} =$~10$^{-6}$ bar) reach the conductivity peak\footnote{We assume that the conductivities do not vary with latitude.} at $\lambda =$~14.5$^{\circ}$.  At this point $B(\lambda,L) =$~1.7~$\times$~10$^{-4}$ T and thus $E_x(L) = -u_y \overline{B}_L \approx$~-170 mV~m$^{-1}$.  The equatorial field strength is $B_{\text{eq}}(L) =$~1.3~$\times$~10$^{-4}$ T and thus the net equatorial electric field at the 10$^{-6}$ bar level is $E_{nx} = u_y (B_{\text{eq}} -  \overline{B}_L) \approx$~-40 mV~m$^{-1}$, which agrees reasonably well with the value of $E_{nx} =$~-30 mV~m$^{-1}$ in Figure~\ref{fig:efields2}.

In this case $E_{nx}$ is negative and thus a downward current flows at the equator with peak $j_x =$~1--2 mA~m$^{-2}$ near the lower boundaries of the M2 and M3 regions\footnote{Note that the direction of the current depends on the orientation of the magnetic field.  The orientation of the field, however, does not affect the direction of the ion drag in our example.}.  The current is strong enough to dominate the neutral momentum balance (Figure~\ref{fig:eq_mom}), and ion drag accelerates the zonal wind in the upper atmosphere until $u_y B_{\text{eq}} \approx u_y \overline{B}_b$ or until a balance with the local pressure gradients is achieved.  The magnitude of the current density $j_x$ is roughly constant with latitude within 20$^{\circ}$ of the equator but beyond 20$^{\circ}$ it slightly decreases with latitude (not shown).  This means that acceleration is most effective at low latitudes, possibly leading to the creation of an equatorial jet in the upper atmosphere instead of constant $u_y$.

The second case demonstrates that ion drag does not in general act as a diffusive (drag) force.  To show this, we assumed that the zonal wind only varies with latitude but remains constant in pressure.  As a result, $E_{nx}$ is positive at the equator and drives an upward current with a peak $j_x =$~3--4 mA~m$^{-2}$.  The polarization field $E_x$ along each magnetic field line depends on the conductivity-weighted mean $[\overline{u_y B}]_L$.  Considering again the magnetic field line with $L =$~1.124, the wind speed at $\lambda =$~14.5$^{\circ}$ is $u_y =$~360 m~s$^{-1}$ and $[\overline{u_y B}]_L \approx$~61 mV~m$^{-1}$.  Thus the estimated $E_{nx} (p_{\text{eq}} = 10^{-6}$ bar$) \approx$~69 mV~m$^{-1}$, which is not too different from the actual value of 52 mV~m$^{-1}$ (Figure~\ref{fig:efields2}).  The positive $E_{nx}$ and upward current arise because the zonal wind speed decreases with latitude and the equatorial magnetic field lines in the upper atmosphere connect to higher latitudes near the boundary of the M1 region (Figure~\ref{fig:geom1}). 

\begin{figure}
  \epsscale{1.35}
  \plotone{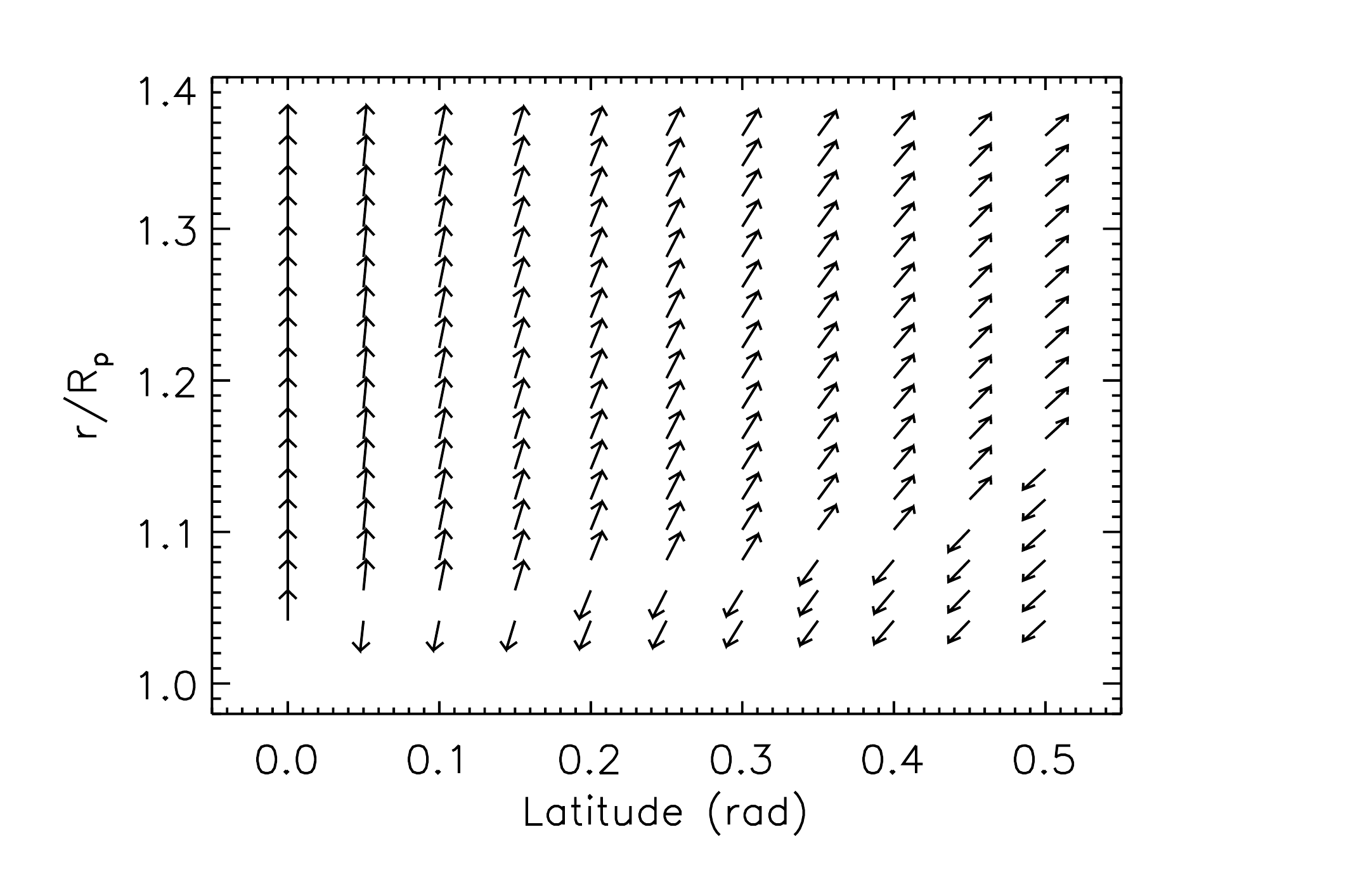}
  \caption{Unit vectors showing the direction of the perpendicular current based on the second example in Section~\ref{subsc:eq_jet} (see text).  The resulting clockwise current loop is closed by field-aligned currents (not shown).}
  \label{fig:exarrow}
\end{figure}

\begin{figure}
  \epsscale{1.15}
  \plotone{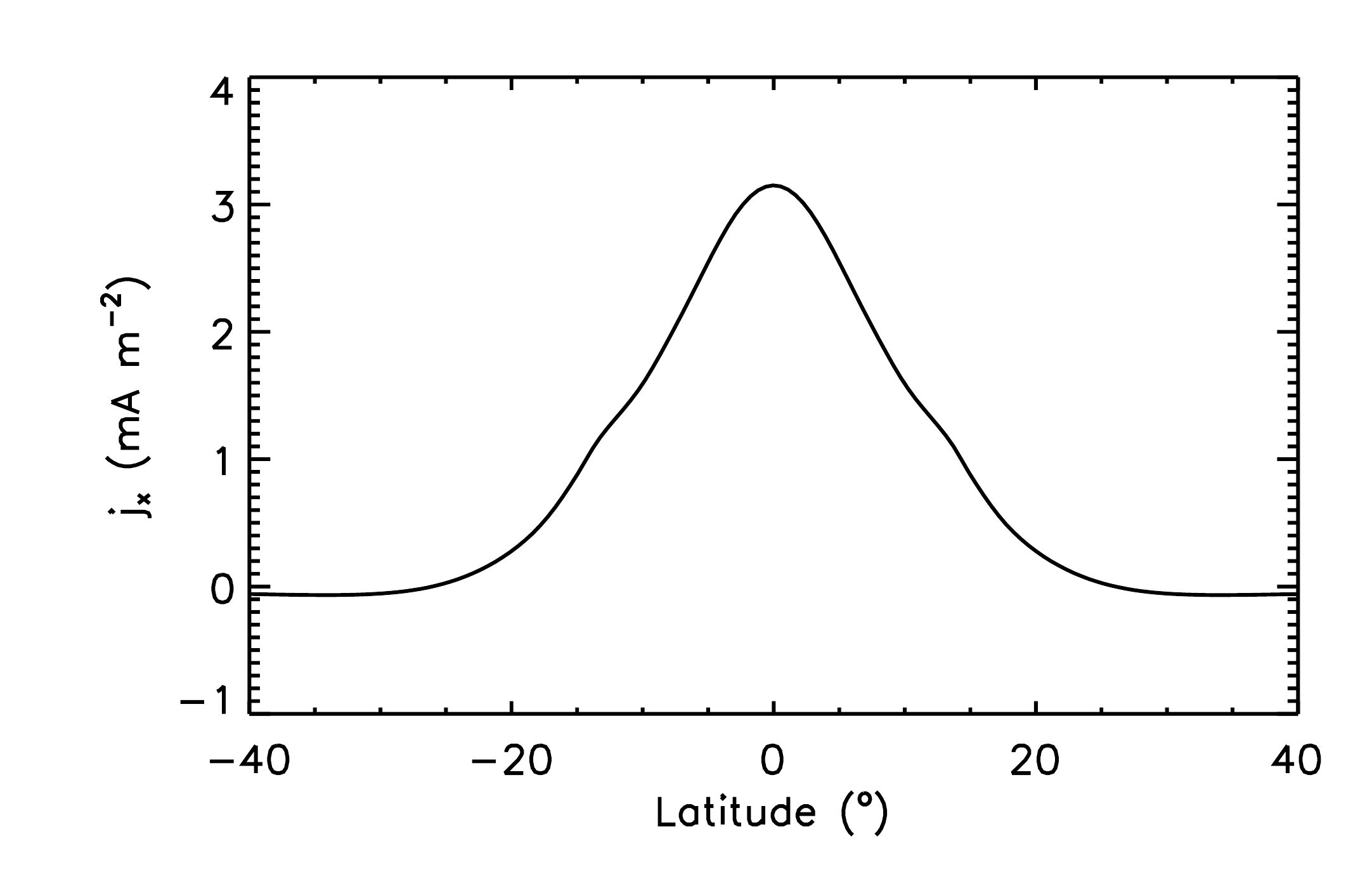}
  \caption{Perpendicular current density $j_x$ as a function of latitude at $p =2 \times$10$^{-6}$ bar.}
  \label{fig:efields3}
\end{figure}

The upward equatorial current decelerates the zonal wind and introduces a vertical gradient into the wind profile.  Figures~\ref{fig:exarrow} and \ref{fig:efields3} illustrate the latitude dependency of the current by showing the direction and magnitude, respectively, of $j_x$ around the equator.  The initial zonal wind profile drives current loops on both sides of the equator.  The current density decreases with latitude and thus ion drag flattens the zonal jet in the upper atmosphere.  The return current near the 10$^{-3}$ bar level is supported by a negative $E_{nx}$ (Figure~\ref{fig:exarrow}) that accelerates the zonal wind in the stratosphere.  Thus ion drag does act like a diffusive drag force at each pressure level within the zonal jet.  This diffusion, however, is limited by the structure of the dipole field and in the vertical direction ion drag introduces gradients into the flow instead of smoothing them.

The third case shows that the pressure profiles predicted by existing close-in exoplanet GCMs for equatorial zonal jets are not necessarily realistic, given the strong influence of ion drag.  The pressure profile of the mean zonal wind shown by Figure~\ref{fig:box_wind} that is motivated by these models leads to the lowest net electric field and current density in the upper atmosphere out of our examples (Figure~\ref{fig:efields2}).  The resulting current loops are similar in morphology to the second example above, but the current density is lower.  Even in this case, however, $l_d$ is much less than $R_p$ in the upper atmosphere and this demonstrates the remarkable sensitivity of the M2 and M3 regions to ion drag.

\subsection{Night side conditions and different magnetic field strengths} 
\label{subsc:magnetic2} 

\begin{figure}
  \epsscale{1.15}
  \plotone{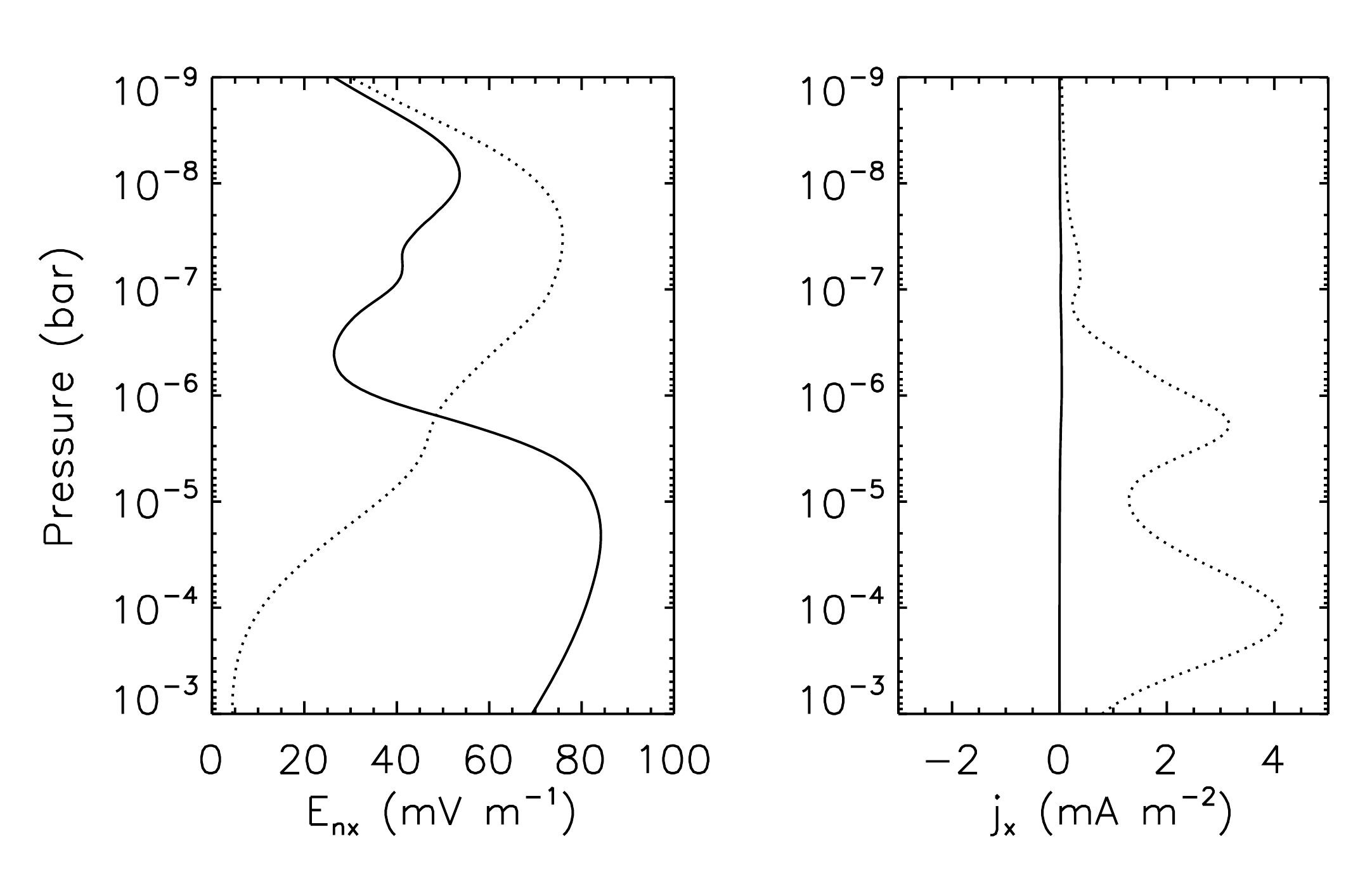}
  \caption{Perpendicular electric field $E_{nx}$ (left panel) and current density $j_x$ (right panel) at the equator based on night side (solid lines) and dayside (dotted lines) conductivities, arising from the equatorial zonal jet that is constant in pressure (Section~\ref{subsc:eq_jet}).  Note that the night side current density ranges from about 10$^{-4}$ to 0.04 mA~m$^{-2}$.}
  \label{fig:exnight}
\end{figure}

As we noted before, night side ion drag is not significant below the 10$^{-6}$ bar level (Section~\ref{subsc:night_cond}).  In order to explore this further, we used the night side Pedersen conductivities (Figure~\ref{fig:night_pedc}) to calculate $E_{nx}$ and $j_x$ based on the equatorial jet that varies with latitude but is constant with pressure (Section~\ref{subsc:eq_jet}).  We find that the night side $E_{nx}$ is not significantly different from the dayside, and in fact higher below the 10$^{-6}$ bar level (Figure~\ref{fig:exnight}).  Because of the lower conductivities, however, the equatorial $j_x$ is substantially lower and as a result $l_d < R_p$ only above the 3~$\times$~10$^{-6}$ bar level.  This agrees with the results in Section~\ref{subsc:night_cond}.  The strong influence of ion drag on the dayside, however, can still modify night side circulation in the middle atmosphere.

Also, in all of the examples (Sections~\ref{subsc:mid_lat_jet} and \ref{subsc:eq_jet}) we assumed that conductivity is constant with latitude and longitude.  While this may be defensible for estimating the Pedersen currents on the dayside, the drop in conductivities in the night side leads to peculiar consequences for the zonal Hall current.  Under the artificial symmetry that we assumed earlier the Hall current is constant with longitude and thus $E_y =$~0.  In the night side, however, the conductivity and thus the Hall current decreases significantly below the 10$^{-6}$ bar level and this should lead to a non-zero $E_y$ that may point from dusk to dawn on the dayside.  Our approach here cannot be used to investigate this or other similar aspects of the complex dynamics between the day and night side that can only be studied properly by circulation models with self-consistent electrodynamics.

In general, the consequences of the reduced conductivity in the night side are somewhat similar to reducing the magnetic field strength.  This is because $l_d$ and magnetization depend on the conductivities, which are reduced either by lowering the electron density or the magnetic field strength.  For example, a dipole moment of $\mu_p =$~0.04 $\mu_J$ is consistent with the lower boundary of the M2 region near the 3~$\times$~10$^{-6}$ bar in the dayside, in which case $l_d < R_p$ only above the 10$^{-6}$ bar level.  We note, however, that the actual boundary of the M2 region in the night side does not depend on the electron density and is relatively insensitive to the changes in the T-P profile.  We can further conclude from this that while there is always potential for strong ion drag in the M2 region, this potential can only be realized if the electron density is sufficiently high.

\begin{figure}
  \epsscale{1.15}
  \plotone{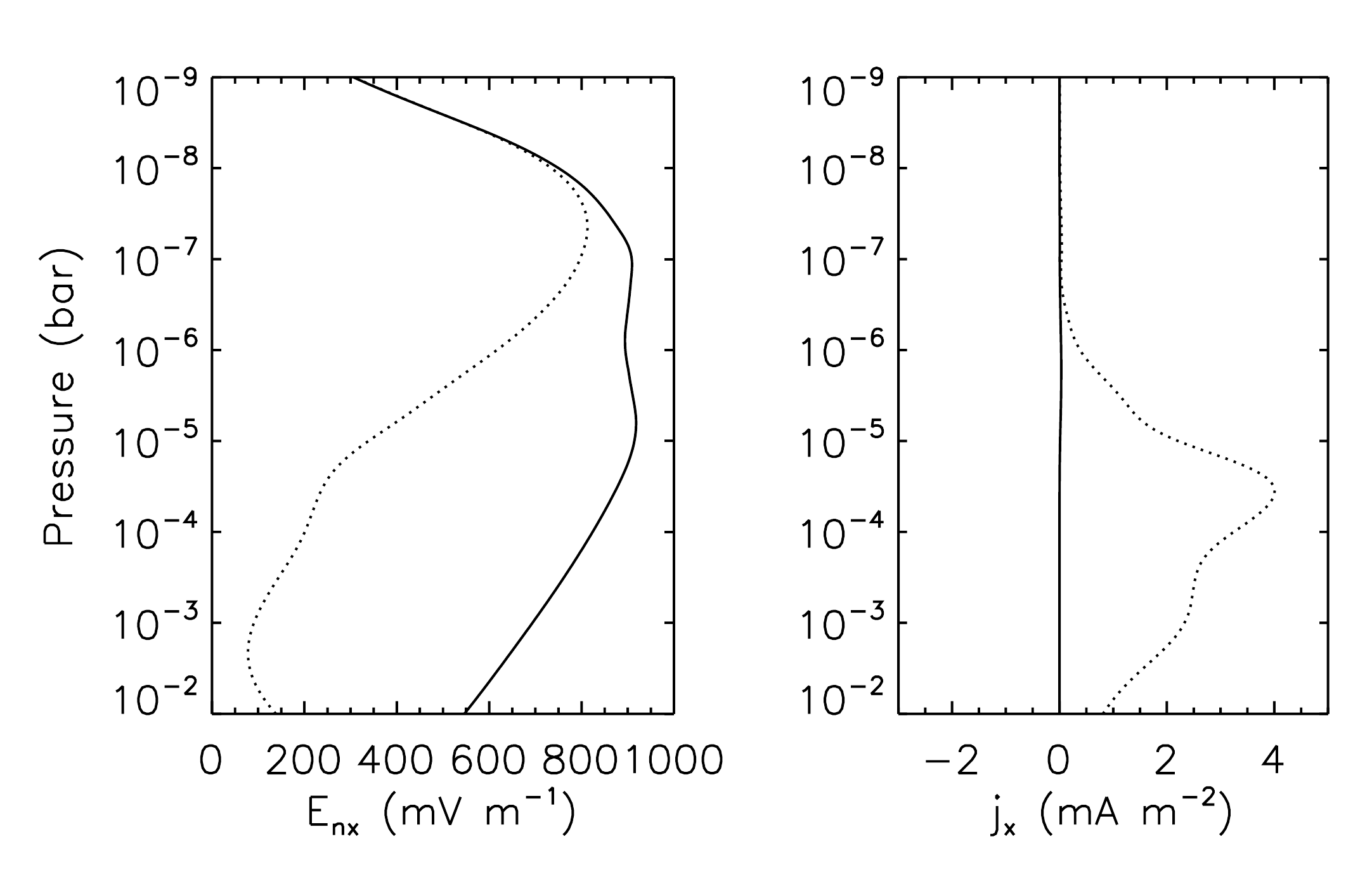}
  \caption{Same as Figure~\ref{fig:exnight}, but with $\mu_p =$~10 $\mu_J$ instead of 1 $\mu_J$.}
  \label{fig:exstrongb}
\end{figure}

A higher magnetic field strength, on the other hand, enhances ion drag in the lower atmosphere, but in a way that is not necessarily straightforward.  As we stated before, $\sigma_P$ decreases in the upper atmosphere with increasing magnetic moment $\mu_p$ as the M2 region moves to higher pressures (Section~\ref{subsc:magnetic}).  For example, if we increase the dipole moment from $\mu_J$ to 10 $\mu_J$, the magnitude of the equatorial $E_{nx}$ increases substantially while the magnitude of the current density $j_x$ does not (Figure~\ref{fig:exstrongb}).  Ion drag, however, also increases with the magnetic field strength through $\mathbf{j} \times \mathbf{B}$ and as a result the region where $l_d < R_p$ shifts down to 3~$\times$~10$^{-3}$ bar on the dayside and 10$^{-5}$ bar in the night side, despite the fact that $j_x$ does not increase substantially.

\subsection{Frictional (resistive) heating}
\label{subsc:joule}


Heating of the neutral atmosphere by electric currents is often mislabeled as `Joule heating' or `Ohmic dissipation'.  In reality, of course, neutrals do not carry charge and there is no reason for them to be directly affected by electric fields or currents.  Instead the neutral atmosphere is heated by mechanical friction that is caused by collisions with the electrons and ions \citep{vasyliunas05}.  By coincidence, the combined heating rate of the plasma and the neutral atmosphere reduces to an expression that has the appearance of conventional Joule heating
\begin{equation}
Q_{J} = \mathbf{j} \cdot \left( \mathbf{E} + \mathbf{u}_n \times \mathbf{B} \right)
\end{equation}
under the same simplifying assumptions that lead to the ion drag term $\mathbf{j} \times \mathbf{B}$ in the momentum equation.  Actual Joule heating or Ohmic dissipation, however, only appears in the combined energy equation of the electrons and ions, and it is typically less significant than frictional heating by collisions.

Estimating the frictional heating rate in actual atmospheres is complicated because the partitioning of ion drag into kinetic and thermal energies cannot be properly determined without self-consistent models of the momentum and energy balance.  In particular, ion drag strongly modifies the dynamics in the upper atmosphere, and the steady state circulation is not known \textit{a priori}.  In the end the frictional heating rate depends on the degree to which the atmosphere is prevented from reaching equilibrium with respect to ion drag by other forces.  More sophisticated models are required to properly address this question in future studies.  Here we concentrate on order of magnitude estimates of $Q_J$ based on the mid-latitude and equatorial jets in Sections~\ref{subsc:mid_lat_jet} and \ref{subsc:eq_jet}, respectively, with the caveat that these estimates may turn out to be substantially different from the actual heating rates. 

\begin{figure}
  \epsscale{1.15}
  \plotone{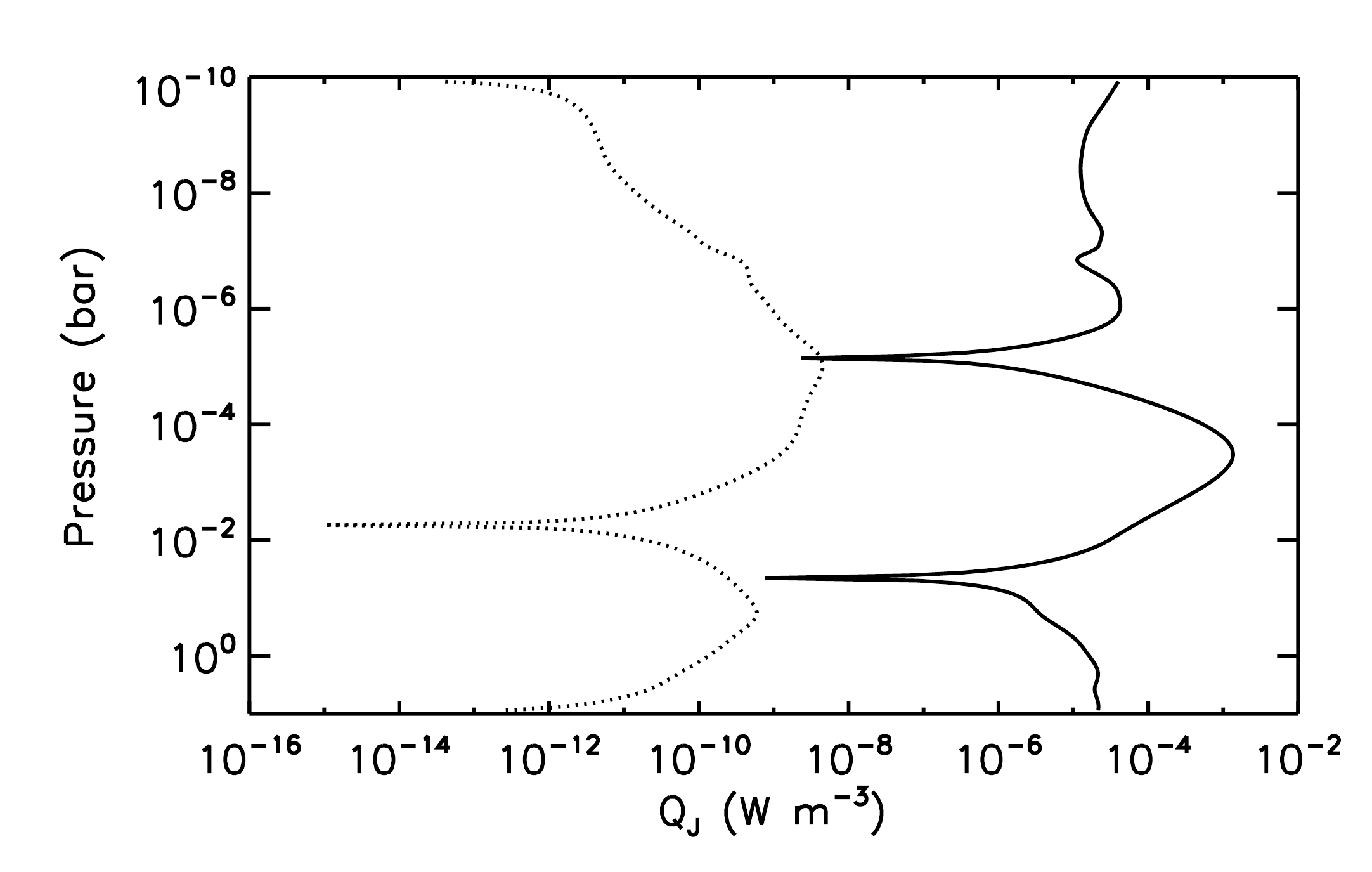}
  \caption{Contributions to the frictional heating rate $Q_J$ from perpendicular [solid line, $j_x (E_x + u_y B_z)$] and field-aligned [dotted line, $j_z E_z$] currents at $\lambda =$~45$^{\circ}$ based on the mid-latitude jet (Figure~\ref{fig:box_wind}).}
  \label{fig:joule1}
\end{figure}

\begin{figure}
  \epsscale{1.15}
  \plotone{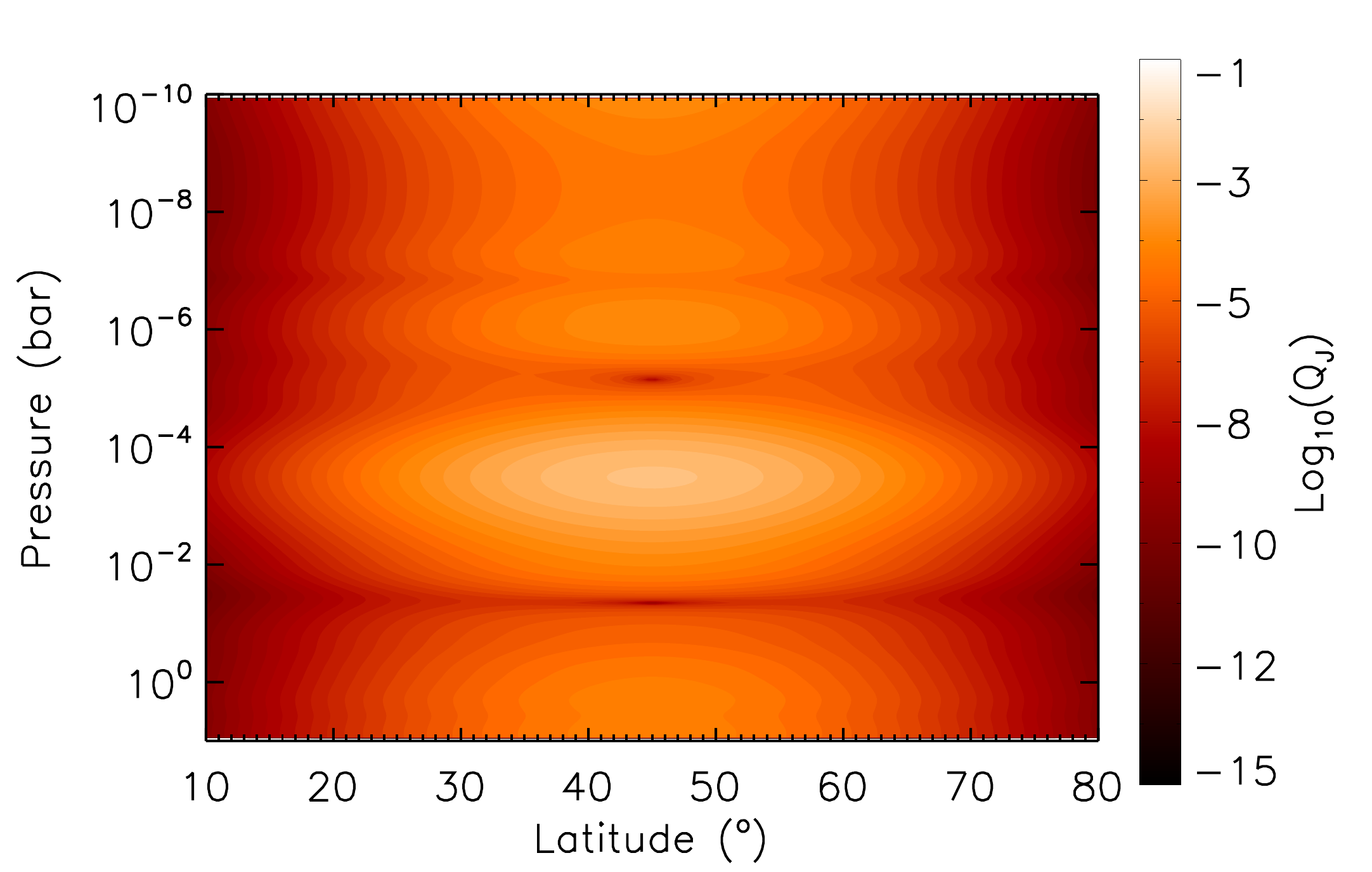}
  \caption{The frictional heating rate $Q_J$ based on the mid-latitude jet in Section~\ref{subsc:mid_lat_jet}.}
  \label{fig:joule2}
\end{figure}

We used the mid-latitude jet to calculate the magnitude of frictional heating, and to explore the relative contribution of the field-aligned and perpendicular currents.  In this case the heating rate is:  
\begin{equation}
Q_J = j_x (E_x + u_y B_0) + j_z E_z.
\end{equation}   
Figure~\ref{fig:joule1} shows the heating rates from field-aligned (vertical) and perpendicular currents at $\lambda =$~45$^{\circ}$ and Figure~\ref{fig:joule2} shows a contour plot of $Q_J$ based on the current loop in Figure~\ref{fig:jx1b}.  These figures show that $Q_J$ peaks near the maximum of the zonal jet, both in pressure and in latitude.  In general, field-aligned currents do not contribute significantly to the heating rate, other than in narrow regions where $j_x$ changes sign. 

In the upper atmosphere $Q_J$ is generally very high.  For example, the bolometric stellar flux at HD209458b is $F_* \approx$~9.8~$\times$~10$^5$ W~m$^{-2}$.  If this flux is absorbed by the atmosphere and distributed globally, the column-integrated frictional heat flux $F_J$ at $\lambda =$~45$^{\circ}$ amounts to about 2.9 \% of the effective stellar flux.  Most of the frictional heating occurs above the 0.01 bar level where the volume heating rate $Q_J$ is comparable or higher than the stellar heating rate.  Remarkably, above the 10$^{-6}$ bar level $F_J$ is more than 1,000 times higher than the stellar X-ray and EUV (XUV) heating rate \citep{koskinen13a,koskinen13b}.  In reality, however, these estimates of $F_J$ are likely to be too high.  This is because the assumed zonal jet will be modified by ion drag above the 10$^{-3}$ bar level.  Nevertheless, the frictional heating rates based on such a jet demonstrate the potential of ion drag to modify the energy balance in the upper atmosphere.

\begin{figure}
  \epsscale{1.15}
  \plotone{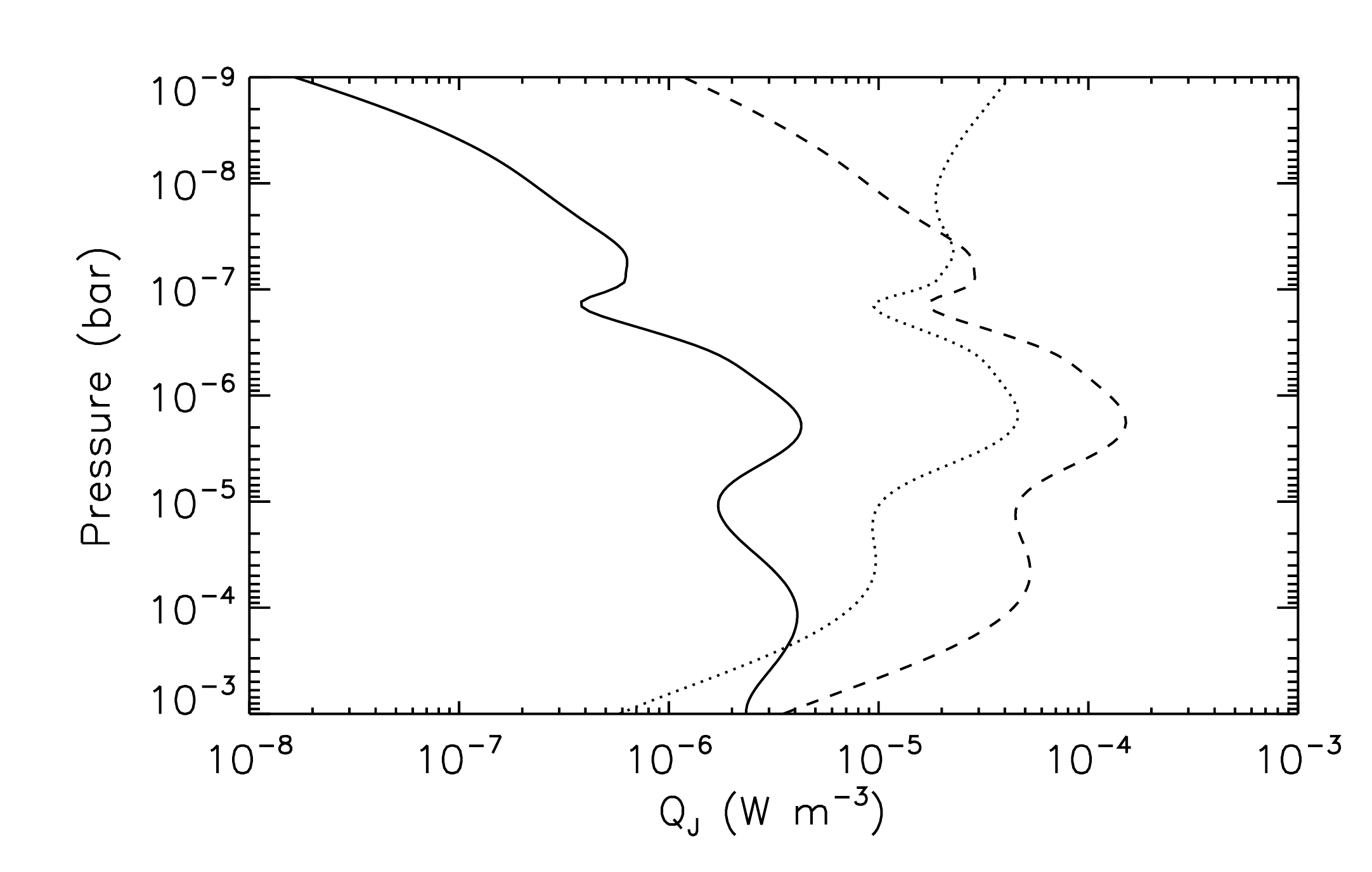}
  \caption{Frictional heating rates at the equator based on currents generated by the three different zonal jets in Section~\ref{subsc:eq_jet}. Results are shown for the realistic zonal jet (solid line), zonal jet that is constant with pressure (dashed line) and a constant zonal wind of 1 km~s$^{-1}$ with both latitude and pressure (dotted line) (see Figure~\ref{fig:efields2}).}
  \label{fig:joule3}
\end{figure}

The influence of the zonal wind profile on the heating rate is further illustrated by Figure~\ref{fig:joule3}, which shows $Q_J$ based on the three different equatorial jets in Section~\ref{subsc:eq_jet}.  The highest $F_J$ above the 10$^{-3}$ bar level, about 5 \% of the stellar flux, is obtained with the constant zonal wind of 1 km~s$^{-1}$ everywhere.  In this case the energy is mostly deposited in the thermosphere where $F_J$ is over 10,000 times higher than the XUV heating rate.  Such a high frictional heating rate is clearly unrealistic, and it demonstrates that the zonal wind will be modified by ion drag instead.  In the second case of a constant zonal wind with pressure, $F_J$ is about 0.5 \% of the stellar flux and still over 1,000 times higher in the thermosphere than the stellar XUV heating rate.  The more realistic mean zonal wind profile in the third case leads to $F_J$ that is about 0.02 \% of the stellar flux and about 30 times higher than the stellar XUV flux in the thermosphere.

\section{Discussion}
\label{sc:discussion}

\subsection{Electron densities}

We have shown how the source of free electrons on close-in EGPs varies both horizontally and with pressure.  In the thermosphere above the 10$^{-6}$ bar level most of the electrons come from photoionization of H on the dayside.  Due to horizontal transport, the long lifetime of H$^+$, and the penetration of the stellar XUV radiation to the night side, the diurnal electron density difference in the thermosphere is at most an order of magnitude \citep{koskinen10}.  We note though that any estimates of the electron density in the thermosphere should be treated with caution.  Basic models of the H/He ionosphere typically overestimate the peak electron density and fail to explain the relatively large diurnal differences observed in the ionospheres of Jupiter and Saturn \citep[e.g.,][]{yelle04b,kliore09}.  These difficulties point to possible deficiencies in our general understanding of H$_2$ physical chemistry \citep[e.g.,][]{hallett05} that can only be resolved by a detailed study of the solar system giant planets.  

A major difference from the solar system giant planets is that alkali metals such as Na and K do not condense on hot close-in EGPs and are present in the atmosphere as atoms.  These metals have a relatively low ionization potential and they are effectively ionized both thermally and by stellar UV radiation.  In a clear atmosphere the relevant UV radiation ($\lambda <$~286 nm) penetrates well past the thermosphere and photoionization dominates over thermal ionization down to the 0.2 bar level \citep{lavvas14}.  As a result, there is a lower ionospheric peak near the 1 mbar level where the electron density is in fact much higher than in the thermosphere and far higher than the electron density in any ionosphere of the solar system.  The recombination time of the alkali metals, however, is relatively short and the electron density in the lower atmosphere, above the 0.2 bar level, reduces by many orders of magnitude in the night side. 

In this regard our results differ from previous studies of ion (magnetic) drag on extrasolar giant planets that have all ignored photoionization and only included electron densities from thermal ionization.  Based on our results for the dayside, the assumption that the atmosphere can be treated as an insulator above the 0.01 bar level \citep[e.g.,][]{batygin10,batygin11} is clearly invalid.  Further, the dayside electron densities are generally much higher than expected from thermal ionization.  Thus currents flow in the upper atmosphere and do not necessarily have to close in the interior.  In this way the prevalence of photoionization in the upper atmosphere significantly complicates electrodynamics on close-in EGPs.  It also reduces the dependency of the atmospheric electron density on temperature only \citep[e.g.,][]{perna12} and makes the conductivities directly dependent on the stellar UV radiation.  

The selection of appropriate heavy elements in calculating the electron densities is also important.  For example, \citet{perna10a,perna10b} included K only and missed an important contribution to the conductivities from the more abundant Na.  \citet{batygin10}, on the other hand, included H, He, Na, K, Li, Rb, Fe, Cs, and Ca while \citet{menou12} included the first 28 elements of the periodic table.  Our calculations show that the inclusion of H, He, Na, K, Mg, and C is sufficient.  The inclusion of these elements is also defensible because, with the exception of He, there is at least some evidence for their presence in the atmospheres of close-in EGPs \citep[e.g.,][]{charbonneau02,vidalmadjar03,fossati10,linsky10,sing11}.  

With regard to the other species, it is particularly important to exclude elements that are likely to condense in the deep atmosphere.  For example, Ca and Fe have strong absorption lines that should be detectable in the observations of HD209458b but are not seen \citep{sing08a,sing08b,lavvas14}.  With the T-P profile for HD209458b, these species are expected to condense below the 10 bar level and thus their lack in the data is probably not a coincidence.  Related to this, none of the heavy elements in our calculations are fully ionized at the pressure levels probed by the observations and thus ionization itself cannot explain the lack of detection.  This, however, means that the non-detection of K on HD209458b is very surprising, and we discuss this problem extensively in a companion paper \citep{lavvas14}.  In general, future observations of metal line absorption on different EGPs can potentially constrain the composition of the ionosphere and thus the role of ion drag in the atmosphere.

\subsection{Conductivities and the Ohm's law}

Electrodynamics is strongly dependent on the electron density and magnetic field strength that vary with pressure.  We used the calculated electron densities to constrain the conductivity regimes on close-in EGPs and provide the appropriate form of the generalized Ohm's law in each regime (Section~\ref{subsc:conds}).  The basic regimes of parallel (Ohmic), Hall, and Cowling (ambipolar) resistivity that depend on the electron and ion magnetization $k_{e,in}$ are well recognized in the solar atmosphere community \citep[e.g.,][]{leake13}, and have recently also been applied to magnetized circumstellar disks \citep[e.g.,][]{bai13}.  The same regimes have also been studied extensively in the terrestrial ionosphere where they are separated based on the parallel, Hall, and Pedersen conductivities instead of resistivity \citep[e.g.,][]{baumjohann97}.  They have not, however, received any attention in the extrasolar planet community until now.      

Models of stellar atmospheres and planetary ionospheres have traditionally relied on two different approaches to electromagnetic effects that, as we have shown, are not as clearly separable on close-in EGPs.  The main differences between the solar chromosphere and planetary ionospheres are the higher ionization fraction and magnetic field strength on the Sun.  As a result, the solar magnetic field is susceptible to instability and perturbations from the plasma motion while perturbations to the magnetic dipole field of the Earth by ionospheric electrodynamics are negligible.  In the latter (static) case the magnetic field is fixed and atmospheric electrodynamics is described in terms of currents and electric fields (the E,j paradigm) while in the former (dynamic) case the magnetic field evolves continuously and magnetohydrodynamics is described in terms of the plasma flow and magnetic field (the B,v paradigm).  In contrast to many astrophysical MHD applications, however, in both cases the conductivity tensor is anisotropic.

The Ohm's law is commonly used in electrodynamic models while the induction equation is used by MHD models.  These equations describe the same physics and they can easily be converted from one to the other by using Faraday's law.  In Section~\ref{subsc:ohms} we present the generalized three-fluid Ohm's law and induction equation in the center of mass rest frame that connect the (E,j) and (B,v) paradigms and are valid for any degree of ionization in the atmosphere.  Such equations are necessary in the atmospheres of close-in EGPs where the ionization fraction and magnetic field strength change with both pressure and horizontal location.  

Previous models of ion drag on EGPs \citep[e.g.,][]{perna10a,perna10b,batygin10} are limited to the lower atmosphere below the 1 mbar level (the M1 region).  In this region the electron/ion gyrofrequencies are much lower than the electron/ion-neutral collision frequencies ($k_{in} << k_{en} <<$~1).  The mechanism that drives the current in the atmosphere (Section~\ref{subsc:m2m3}) relies on the right combination of gyromotion and collisions, and once $k_{en} <<$~1 this mechanism is largely suppressed.  Further, in our case the plasma frequency $\Omega_e << \nu_{en}$ below the 0.01 bar level on the dayside and below the 10$^{-4}$ bar level in the night side.  Thus plasma behavior is suppressed in the lower atmosphere and both electrons and ions simply equilibrate with the neutral atmosphere \citep[e.g,][]{baumjohann97}.  

Our derivation of the Ohm's law, however, appears to contradict this simple physical intuition.  In particular, in the M1 region the generalized Ohm's law reduces to equation~(\ref{eqn:perna}) with isotropic conductivity $\sigma_{en}$ that has been used in practically all of the previous exoplanet studies.  There is no reason to doubt the general validity of this equation in the M1 region.  In agreement with our basic physical intuition, however, the neutral momentum equation and the Ohm's law can be used to show that ion drag is not important in the M1 region.  Physically speaking, the length scale $l_d$ (Section~\ref{sc:emdyn}) over which the current (or ion drag) becomes significant increases rapidly as $k_{en}$ decreases.  As a general rule, $l_d >> R_p$ in the M1 region, even if one assumes the maximum current generation by neutral winds with zero polarization electric fields.  

It can be shown that $l_d$ is related to the Lundquist length scale, which is the length scale over which the Lundquist number $S \approx$~1.  The Lundquist number plays the same role in predominantly neutral atmospheres as the magnetic Reynolds number $R_m$ plays in fully ionized plasmas by quantifying the degree to which the background magnetic field is perturbed by plasma motions.  In contrast to the standard $R_m$ \citep[e.g.,][]{baumjohann97}, the Lundquist number accounts for plasma-neutral collisions and the stability of the magnetic dipole field that is continuously regenerated in the deep interior of the planet.  Our results show that $S <<$~1 in the M1 region, indicating that perturbations to the dipole field due to plasma motions in the lower atmosphere should be small. 

Interestingly, our results provide some justification for studies that use equation~(\ref{eqn:perna}) in steady state to estimate frictional heating rates in the deep atmosphere based on predetermined wind patterns \citep[e.g.,][]{batygin10}.  This approach is technically acceptable because ion drag is negligible and the magnetic field should be largely unperturbed by dynamics in the atmosphere -- although it is limited by the neglect of the interior currents that give rise to the planetary magnetic field and currents in the upper atmosphere.  From a physical perspective, there is always some degree of friction, no matter how small, between the neutral atmosphere and the plasma that attempts to separate from the neutrals.    

Our treatment of the electrodynamics in the M2 and M3 regions of close-in EGPs is entirely new.  In these regions the conductivity tensor becomes anisotropic, with the degree of anisotropy increasing with the magnetic field strength.  In the M2 region the electrons partly decouple from the neutrals while ions remain coupled to the neutrals ($k_{en} >$~1, $k_{in} <$~1) and in the M3 region both ions and electrons decouple from the neutrals ($k_{en} > k_{in} >$~1).  The parallel (magnetic field-aligned) conductivity in the M2 and M3 regions is typically much larger than the perpendicular Hall and Pedersen conductivities.  By design, the Hall conductivity is always higher than the Pedersen conductivity in the M2 region while the Pedersen conductivity is much higher than the Hall conductivity in the M3 region. 

The partial decoupling of the plasma from the neutral atmosphere promotes current generation over short length scales (Section~\ref{subsc:m2m3}) and thus $l_d << R_p$ on the dayside above the 1 mbar level (Section~\ref{sc:emdyn}).  With magnetic moments close to $\mu_J$ or higher, ion drag is so strong that it is likely to dominate the momentum balance in the M2 and M3 regions together with the pressure gradients arising from stellar irradiation.  In addition, the Lundquist number approaches and exceeds unity in the upper atmosphere, implying that the dipole magnetic field is susceptible to perturbations by the currents in the atmosphere.

The short recombination time of the alkali metals, however, means that the influence of ion drag is much less prominent in the night side.  Effectively, the reduced conductivity in the night side has the same effect as lowering the magnetic field strength on the dayside by a factor of 10--100.  This leads to an interesting dichotomy between the dynamics in the dayside and the night side.  Given that large scale dynamics on different exoplanets can potentially be constrained by observations \citep[e.g.,][]{knutson07,snellen10,showman13}, this dichotomy is worthy of further study by MHD models with anisotropic conductivity that can capture the turbulent transition across the terminator.

In contrast to the conductivities, magnetization does not change significantly from the dayside to the night side.  This means that while the M2 and M3 regions are always potentially affected by ion drag, this potential is only realized if the conductivities are sufficiently high.  Obviously, magnetization depends on the assumed magnetic field strength.  With a higher magnetic moment of 10 $\mu_J$, the M2 region extends down to the 0.01 bar level, thus also extending the region of anisotropic conductivity deeper into the atmosphere.  With a magnetic moment of 0.1 $\mu_J$, on the other hand, ion drag is only important in the thermosphere even on the dayside.            

Finally, in the M4 region above the 10$^{-8}$ bar level the ionization fraction is sufficiently high ($x_e \gtrsim 10^{-3}$) for the plasma to become completely decoupled from the neutrals.  In this regime the Ohm's law in the plasma frame (equation~\ref{eqn:ion_ohm}) reduces to the fully ionized limit.  This region is important because on close-in EGPs it is believed to be escaping hydrodynamically, with detectable signatures in UV transit observations \citep{vidalmadjar03,vidalmadjar04,fossati10,linsky10,lecavelier12}.  We find that ion drag on the neutral atmosphere in the upper atmosphere is relatively strong, raising the possibility that the capture of the plasma on closed field lines can also affect neutral mass loss rates \citep{adams11,trammell11}.

The combined thermal and dynamical pressure of the upper atmosphere, however, exceeds the magnetic pressure at the equator for $\mu_p \lesssim 0.5 \mu_J$ and this opens the possibility that the magnetic field lines follow the escaping plasma, in the same manner as they do in the solar wind.  Further, even if ion drag is important in the upper atmosphere, it is not clear to what degree it inhibits mass loss from close-in EGPs as the capture of the escaping gas on closed field lines leads to significant heating of the upper atmosphere that can possibly restore the escape rate \citep{yelle04}.  The development of a two fluid model to simulate the plasma and the neutral atmosphere with the induction equation for the magnetic field is a useful future avenue to addressing these questions.

\subsection{Electrodynamics and frictional heating}  

In order to better understand the basic mechanisms of electrodynamics, we studied highly idealized mid-latitude and equatorial zonal jets.  These jets have a Gaussian shape in latitude and a pressure dependency that is designed to mimic the zonal mean circulation on close-in EGPs such as HD209458b \citep{showman09,koskinen10}.  We relied on the static approximation (equation~\ref{eqn:estatic1}) although, as we have explained, this approximation needs to be revised in future work.  In contrast to some of the previous studies \citep[e.g.,][]{menou12,rauscher13}, we showed that ion drag does not always act like a diffusive drag force and it cannot be modeled as Rayleigh drag.  Instead, we find that ion drag attempts to eliminate variations of ($-\mathbf{u}_n \times \mathbf{B}$) along the magnetic field lines and in many cases this can lead to the acceleration of the winds and the introduction of both vertical and meridional gradients into the flow.  We also showed that ion drag does not generally vanish at the equator.  

At mid-latitudes the dipole magnetic field lines are nearly vertical and penetrate practically through the whole atmosphere.  Thus the consequences of anisotropic conductivity in the M2 and M3 regions are also felt in the M1 region, and the polarization electric field is roughly constant and equal to the conductivity-weighted mean of $\mathbf{u}_n \times \mathbf{B}$ along the magnetic field lines (Section~\ref{subsc:mid_lat_jet}).  With a constant magnetic field, ion drag attempts to remove variations of the wind speed along the field lines.  The degree to which this is possible depends on the strength of the ion drag forces compared to the other terms in the momentum equation.  Because ion drag is much less important below the 1 mbar level than it is in the upper atmosphere, circulation in the lower atmosphere can drive the dynamics at higher altitudes.   

At the equator the horizontal magnetic field lines with L ($=r_{\text{eq}}/R_p$) values in the M2 and M3 regions connect to relatively low latitudes.  There is only a narrow region in the equatorial M1 region where the field lines do not pass through the M2 and M3 regions that we ignore in this work.  The detailed structure of the currents in the upper atmosphere arising from the equatorial jet depends sensitively on the assumed vertical and meridional wind profiles but typically symmetric loops appear on both sides of the equator (Section~\ref{subsc:eq_jet}).  We find that zonal winds predicted by current exoplanet GCMs are likely to be significantly modified above the 1 mbar level on planets where the magnetic moment is comparable to $\mu_J$.  We also find that the magnetic dipole field geometry alone drives significant ion drag in the upper atmosphere even when the zonal wind is initially constant with altitude and latitude.

We note that the zonal jets give rise to ion drag on the meridional and vertical winds by a zonal Hall current that we ignore in this work.  This current does not contribute to frictional heating because under the (artificial) symmetry of our examples $E_y =$~0 and the Hall current is perpendicular to the net electric field in the $x$ direction.  In reality the Hall current in the M2 region cannot be constant in longitude because of the large drop in conductivity in the night side.  This means that a non-zero zonal electric field with either dusk to dawn or dawn to dusk orientation can develop on the dayside, depending on the direction of the zonal current (that can vary with pressure).  This complication should be studied by more comprehensive models in the future that can properly address the diurnal differences on close-in EGPs.

Finally, ion drag heats the atmosphere mostly through mechanical friction from plasma-neutral collisions \citep{vasyliunas05}.  We show that the frictional heating rate is strongly dependent on the assumed vertical and horizontal wind profile.  Whether or not the currents contribute to the kinetic energy of the atmosphere through ion drag or to the thermal energy through friction depends on the momentum balance that needs to be worked out by self-consistent models.  Nevertheless, our results demonstrate the potential for significant frictional heating in the upper atmosphere.

The column-integrated frictional heat flux $F_J$ based on the mid-latitude jet in Section~\ref{subsc:mid_lat_jet} amounts to about 2.9 \% of the globally averaged stellar bolometric flux.  The different equatorial jets in Section~\ref{subsc:eq_jet} produce values of $F_J$ between 0.02 and 5 \%.  These heat fluxes on the dayside are small but not negligible, and if they are deposited deep enough they could affect the radius evolution of the planet \citep{batygin11}.  Our calculations at the equator are limited to the upper atmosphere above the 1 mbar level, but the mid-latitude solution extends to 10 bar.  In this latter case we find that most of the frictional heat is actually deposited in the high altitude current loop above the 0.01 bar level.  The largest relative effect from frictional heating is felt in the thermosphere where the heat flux is 10--10,000 times higher than the stellar XUV heating rate, depending on the assumed zonal wind profile.

Clearly, some of these heat fluxes are unrealistically high due to the lack of self-consistent dynamics.  Nevertheless, frictional heating of the upper atmosphere by electric currents can be important because it effectively uses the energy from circulation that is powered by stellar insolation at lower altitudes.  Even a small fraction of the bolometric flux that heats the lower atmosphere is large compared to the stellar XUV flux.  In the thermosphere, additional heating can broaden the absorption lines of atoms and ions and affect the interpretation of UV transit observations \citep{benjaffel10,koskinen10b}.  Frictional heating from ion drag can also enhance the energy-limited mass loss rate from close-in EGPs.  In the solar system, ion drag can help to explain the abnormally high temperatures in the thermospheres of the giant planets \citep{smith13}.  Self-consistent models of ion drag on giant planets, both in the solar system and extrasolar planets, are therefore an interesting direction for future research.            

\section{Conclusions}   
\label{sc:conclusions}

The aim of this work was to use electron densities calculated by a new photochemical model \citep{lavvas14} to constrain electrodynamics and ion drag in the atmospheres of close-in EGPs.  The electron densities are based on a solar composition of atoms such as H, Mg, Na and K that have been detected in Hot Jupiter atmospheres \citep[e.g.,][]{vidalmadjar03,fossati10,charbonneau02,sing11}.  We used the properties of the transiting planet HD209458b in these calculations, although the results have general validity and are not necessarily intended to represent any specific planet.  We showed that close-in EGPs have ionospheres that extend to the lower atmosphere with electron densities that are much higher than the electron densities in any ionosphere of the solar system.  As a result, the electrical conductivity in the upper atmospheres of close-in EGPs can be comparable to the corresponding conductivity in the solar chromosphere.   

In line with \citet{koskinen10}, we find that photoionization of H and He dominates in the thermosphere above the 10$^{-6}$ bar level.  In this region diurnal variations in electron density are also relatively small.  Photoionization of metals such as Mg, Na, and K, however, creates a lower ionospheric peak near the 1 mbar level in the dayside where the electron density is higher than in the thermosphere.  Photoionization dominates over thermal ionization on the dayside down to about 0.2 bar while thermal ionization produces most of the free electrons at deeper pressures.  As a result, future observations of species like Mg, Na, and K can be used to characterize the ionosphere further, particularly if their abundances can be better constrained.  Previous studies of ion drag on close-in EGPs ignored photoionization \citep{perna10a,perna10b,batygin10} and thus severely underestimated the dayside electron densities.   

Phase curve observations of Hot Jupiters \citep[e.g.,][]{knutson07} can potentially be used to constrain the effect of ion drag on global dynamics.  Due to the neglect of photo-ionization, however, previous models have underestimated the day-night contrast in conductivity in the middle atmosphere.  They have also relied on Rayleigh drag to simulate the global effects of ion drag.  In our examples ion drag attempts to eliminate variations of ($-\mathbf{u}_n \times \mathbf{B}$) along the magnetic field lines.  Sometimes this leads to behavior that is similar to a frictional drag force -- at other times ion drag actually introduces vertical and horizontal structure to the wind profiles.  We also find that ion drag does not vanish at the equator \citep[e.g.,][]{menou12}.  In fact vertical currents that lead to ion drag at the equator are generally required to close the current loops generated by an equatorial zonal jet.  In light of these findings, we feel that the consequences of ion drag on global dynamics may have to be re-evaluated.  

Another potentially interesting consequence of photo-ionization is that it makes the conductivities and thus ion drag and resistive heating dependent on the stellar UV fluxes rather than temperature only.  This raises the possibility of correlations between the stellar UV output and observable properties of transiting planets such as brightness temperatures.  These correlations might be difficult to recognize because the atmospheric temperatures obviously also depend on the stellar flux, and much of the UV radiation responsible for ionizing, say, Na and K is in the blackbody continuum.  Nevertheless, both FUV radiation and X-rays, that do not necessarily correlate strongly with the stellar continuum, penetrate to the middle atmosphere and could form the basis for looking for such correlations in the future. 

In addition, our results point to particularly strong effects of ion drag in the upper atmosphere.  We divided the atmosphere into four different regimes (M1, M2, M3, and M4) based on conductivities and magnetization, and provided the three-fluid non-ideal MHD equations that are valid in these regimes.  Previous models have not included the atmosphere above the 1 mbar level (the M2, M3, and M4 regions).  In these regions the electrons and ions partly decouple from the neutral atmospheres and the combination of gyromotion and less frequent collisions leads to charge separation and strong currents.  We demonstrate that, with the Jovian magnetic moment, currents and ion drag in the middle and upper atmosphere are sufficiently strong to dominate the momentum balance, together with the neutral pressure gradients that arise from stellar insolation and subsequent dynamics.  

Ion drag also affects the thermal energy balance of the atmosphere mostly through frictional heating arising from plasma-neutral collisions \citep{vasyliunas05}.  Based on our examples, we find column-integrated local maximum heat fluxes that amount to 0.4--5\% of the globally averaged stellar bolometric flux, deposited mostly in the upper atmosphere above the 0.01 bar level.  The potential for frictional heating is particularly strong in the thermosphere where ion drag may enhance mass loss rates and affect the interpretation of UV transit observations \citep[e.g.,][]{benjaffel10,koskinen13b}.  A magnetic field can also affect the morphology of the escaping plasma around planets like HD209458b, and this opens the possibility for future observations of escaping atmospheres as a means to place some constraints on the magnetic field strengths, even though the interpretation of such observations is likely to remain ambiguous for a long time to come.

Ion drag and frictional heating may also prove to be important on the giant planets in the solar system.  A recent study by \citet{smith13} shows in principle that heating arising from electrodynamic coupling of Jupiter's thermosphere and stratosphere can help to explain the relatively high temperatures in the Jovian thermosphere.  In addition to direct heating, other studies on Saturn raise the question of whether wind-driven ion drag could change the predicted circulation and help to redistribute energy from the polar auroral region to low and middle latitudes more efficiently \citep[e.g.,][]{mullerwodarg12,koskinen13c}.  In the absence of direct constraints on the winds in the upper atmosphere, however, these questions can only be properly addressed with self-consistent models of the momentum and energy balance.                                                   

\acknowledgments

TTK and RVY acknowledge support by the National Science Foundation (NSF) grant AST 1211514.  JYKC acknowledges the hospitality of the Kavli Institute for Theoretical Physics (KITP), Santa Barbara.  PL acknowledges financial support from the Programme National de Plan\'etologie (PNP).  We thank J. Erwin, R. Jokipii, J. Giacalone, and V. Vasyliunas for useful discussions and correspondence.         



\begin{thebibliography}{}
\bibitem[Adams(2011)]{adams11} Adams, F. C. 2011, ApJ, 730, 27
\bibitem[Bai et al.(2013)]{bai13} Bai, X-N., \& Stone, J. M. 2013,  ApJ, 769, 76  
\bibitem[Batalha et al.(2013)]{batalha13} Batalha, N. M., et al. 2013, ApJS, 204, 24
\bibitem[Batygin and Stevenson(2010)]{batygin10} Batygin, K., \& Stevenson, D. J. 2010,  ApJL, 714, 238
\bibitem[Batygin et al.(2011)]{batygin11} Batygin, K., Stevenson, D. J., \& Bodenheimer, P. H. 2011, ApJ, 738, 1
\bibitem[Baumjohann and Treumann(1997)]{baumjohann97} Baumjohann, W., \& Treumann, R. A. 1997, Basic space plasma physics.  (London, England: Imperial College Press)
\bibitem[Ben-Jaffel and Hosseini(2010)]{benjaffel10} Ben-Jaffel, L., \& Hosseini, S. S. 2010, ApJ, 709, 1284
\bibitem[Ben-Jaffel and Ballester(2013)]{benjaffel13} Ben-Jaffel, L., \& Ballester, G. E. 2013, A\&A, 553, A52
\bibitem[Charbonneau et al.(2002)]{charbonneau02} Charbonneau, D., Brown, T. M., Noyes, R. W., \& Gilliland, R. L. 2002, ApJ, 568, 377
\bibitem[Cho (2008)]{cho08} Cho, J. Y-K. 2008, Phil. Trans. R. Soc. A, 366, 4477
\bibitem[Fossati et al.(2010)]{fossati10} Fossati, L., et al. 2010, ApJL, 714, 222
\bibitem[Garcia Munoz(2007)]{garciamunoz07} Garcia Munoz, A. 2007, Plan. Space Sci., 55, 1426
\bibitem[Grie\ss meier et al.(2004)]{griesmeier04} Grie\ss meier, J.-M., et al. 2004,  A \& A, 425, 753
\bibitem[Guillot et al.(1996)]{guillot96} Guillot, T., Burrows, A., Hubbard, W. B., Lunine, J. I., \& Saumon, D. 1996, ApJL, 459, L35
\bibitem[Hallett et al.(2005)]{hallett05} Hallett, J. T., Shemansky, D. E., \& Liu, X. 2005, ApJ, 624, 448
\bibitem[Kliore et al.(2009)]{kliore09} Kliore, A. J., et al. 2009, J. Geophys. Res., 114,  A04315
\bibitem[Knutson et al.(2007)]{knutson07} Knutson, H. A., et al. 2007, Nature, 447, 183
\bibitem[Koskinen et al.(2010a)]{koskinen10} Koskinen, T. T., Cho, J. Y-K., Achilleos, N., \& Aylward, A. D. 2010a, ApJ, 722, 178
\bibitem[Koskinen et al.(2010b)]{koskinen10b} Koskinen, T. T., Yelle, R. V., Lavvas, P., \& Lewis, N. 2010b, ApJ, 723, 116
\bibitem[Koskinen et al.(2013a)]{koskinen13a} Koskinen, T. T., Harris, M., Yelle, R. V., \& Lavvas, P. 2013a, Icarus, 226, 1678
\bibitem[Koskinen et al.(2013b)]{koskinen13b} Koskinen, T. T., Harris, M., Yelle, R. V., \& Lavvas, P. 2013b, Icarus, 226, 1695
\bibitem[Koskinen et al.(2013c)]{koskinen13c} Koskinen, T. T., et al. 2013c, Icarus, 226, 1318
\bibitem[Lavvas et al.(2014)]{lavvas14} Lavvas, P., et al. 2014, submitted to ApJ
\bibitem[Leake et al.(2013)]{leake13} Leake, J. E., et al. 2013, arXiv:1310.0405v2
\bibitem[Lecavelier des Etangs et al.(2012)]{lecavelier12} Lecavelier des Etangs, A., et al. 2012, A \& A, 543, L4
\bibitem[Linsky et al.(2010)]{linsky10} Linsky, J. L., et al. 2010, ApJ, 717, 1291
\bibitem[Menou(2012)]{menou12} Menou, K. 2012, ApJ, 745, 138
\bibitem[Moses et al.(2011)]{moses11} Moses, J. I., et al. 2011, Astrophys. J., 737, 15.
\bibitem[M\"uller-Wodarg et al.(2012)]{mullerwodarg12} M\"uller-Wodarg, I. C. F., Moore, L., Galand, M., Miller, S., \& Mendillo, M. 2012, Icarus, 221, 481
\bibitem[Perna et al.(2010a)]{perna10a} Perna, R., Menou, K., \& Rauscher, E. 2010a, ApJ, 719, 1421
\bibitem[Perna et al.(2010b)]{perna10b} Perna, R., Menou, K., \& Rauscher, E. 2010b, ApJ, 724, 313
\bibitem[Perna et al.(2012)]{perna12} Perna, R., Heng, K., \& Pont, F. 2012, ApJ, 751, 59
\bibitem[Rauscher and Menou(2013)]{rauscher13} Rauscher, E., \& Menou, K. 2013, ApJ, 764, 103 
\bibitem[Richmond(1995)]{richmond95} Richmond, A. D. 1995, J. Atmos. Terr. Phys., 57, 1103
\bibitem[Richmond and Thayer(2000)]{richmond00} Richmond, A. D., \& Thayer, J. P. 2000, in Magnetospheric Current Systems, Geophysical Monograph (American Geophysical Union), 118, 131
\bibitem[Rogers and Showman(2014)]{rogers14} Rogers, T. M., \& Showman, A. P. 2014, ApJL, 782, 4
\bibitem[Salby(1996)]{salby96} Salby, M. L. 1996, Fundamentals of atmospheric physics (San Diego, CA, USA: Academic Press)
\bibitem[Schunk and Nagy(2000)]{schunk00} Schunk, R. W., \& Nagy, A. F. 2000, Ionospheres: Physics, plasma physics, and chemistry (Cambridge, England: Cambridge University Press)
\bibitem[Sing et al.(2008a)]{sing08a} Sing, D. K., Vidal-Madjar, A., Desert, J.-M., Lecavelier des Etangs, A., \& Ballester, G. 2008a, ApJ, 686, 658
\bibitem[Sing et al.(2008b)]{sing08b} Sing, D. K., et al. 2008b, ApJ, 686, 667
\bibitem[Sing et al.(2011)]{sing11} Sing, D. K., et al. 2011, A \& A, 527, A73
\bibitem[Showman and Guillot(2002)]{showman02} Showman, A. P., \& Guillot, T. 2002, A \& A, 385, 166
\bibitem[Showman et al.(2009)]{showman09} Showman, A. P., et al. 2009, ApJ, 699, 564
\bibitem[Showman et al.(2013)]{showman13} Showman, A. P., Fortney, J. J., Lewis, N. K., \& Shabram, M. 2013, ApJ, 762, 24
\bibitem[Smith(2013)]{smith13} Smith, C. G. A. 2013, Icarus, 226, 923
\bibitem[Snellen et al.(2010)]{snellen10} Snellen, I. A. G., de Kok, R. J., de Mooij, E. J. W., \& Albrecht, S. 2010, Nature, 465, 1049
\bibitem[Tenenbaum et al.(2014)]{tenenbaum14} Tenenbaum, P., et al. 2014, ApJS, 211, 6
\bibitem[Trammell et al.(2011)]{trammell11} Trammell, G. B., Arras, P., \& Li, Z.-Y. 2011, ApJ, 728, 152
\bibitem[Udry and Santos(2007)]{udry07} Udry, S., \& Santos, N. C. 2007, Annu. Rev. Astron. Astrophys., 45, 397
\bibitem[Vasyli$\overline{\text{u}}$nas and Song(2005)]{vasyliunas05} Vasyli$\overline{\text{u}}$nas, V. M., \& Song, P. 2005, J. Geophys. Res., 110, A02301
\bibitem[Vidal-Madjar et al.(2003)]{vidalmadjar03} Vidal-Madjar, A., et al. 2003, Nature, 422, 143
\bibitem[Vidal-Madjar et al.(2004)]{vidalmadjar04} Vidal-Madjar, A., et al. 2004, ApJL, 604, 69
\bibitem[Wardle(2007)]{wardle07} Wardle, M. 2007, Astrophys. Space Sci., 311, 35
\bibitem[Yelle(2004)]{yelle04} Yelle, R. V. 2004, Icarus, 170, 167
\bibitem[Yelle and Miller(2004)]{yelle04b} Yelle, R. V., \& Miller, S. 2004, in Jupiter. The Planet, Satellites and Magnetosphere (Cambridge, England: Cambridge University Press), 185  
\end{thebibliography}
\end{document}